\documentclass[11pt,a4paper]{article}

\usepackage[dvipdfmx]{graphicx}
\usepackage{jcappub}
\usepackage{color}
\usepackage{bm,latexsym,amsmath,amssymb,amsfonts,mathrsfs,ulem}

\newcommand{\ba}{\begin{eqnarray}}
\newcommand{\ea}{\end{eqnarray}}
\newcommand{\baa}{\begin{align}}
\newcommand{\eaa}{\end{align}}

\newcommand{\simgt}{\lower.5ex\hbox{$\; \buildrel > \over \sim \;$}}
\newcommand{\simlt}{\lower.5ex\hbox{$\; \buildrel < \over \sim \;$}}
\newcommand*{\eps}{\varepsilon}
\newcommand*{\Lag}{{\cal L}}

\newcommand*{\Ad}{{\dot A}}
\newcommand*{\As}{A_*}
\newcommand*{\Asd}{{\dot A}_*}
\newcommand*{\Ah}{{\hat A}}
\newcommand*{\Ahd}{{\dot {\hat A}}}
\newcommand*{\A}{{\cal A}}
\newcommand*{\B}{{\cal B}}
\newcommand*{\C}{{\cal C}}
\newcommand*{\D}{{\cal D}}
\newcommand*{\E}{{\cal E}}
\newcommand*{\F}{{\cal F}}
\newcommand*{\M}{{\cal M}}

\def\({\Bigl(}
\def\){\Bigr)}
\def\[{\Biggl[}
\def\]{\Biggr]}

\title{Extended vector-tensor theories}

\author{Rampei Kimura,}
\emailAdd{rampei@th.phys.titech.ac.jp}
\author{Atsushi Naruko,}
\emailAdd{naruko@th.phys.titech.ac.jp}
\author{Daisuke Yoshida}
\emailAdd{yoshida@th.phys.titech.ac.jp}

\affiliation{
Department of Physics, Tokyo Institute of Technology,
2-12-1 Ookayama, Meguro-ku, Tokyo 152-8551, Japan
	}

\abstract{Recently, several extensions of massive vector theory in curved space-time have been proposed in many literatures. In this paper, we consider the most general vector-tensor theories that contain up to two derivatives with respect to metric and vector field.~By imposing a degeneracy condition of the Lagrangian~in the context of ADM decomposition of space-time to eliminate an unwanted mode,~we construct a new class of massive vector theories where five degrees of freedom can propagate, corresponding to three for massive vector modes and two for massless tensor modes.
We find that the generalized Proca and the beyond generalized Proca theories up to the quartic Lagrangian, which should be included in this formulation, are 
degenerate theories even in curved space-time. Finally, introducing new metric and vector field transformations, we investigate the properties of thus obtained theories under such transformations.
}
\begin{document}
\maketitle

\section{Introduction}

The late-time accelerating expansion of the universe \cite{Perlmutter:aa,Riess:aa} is one of the most challenging and intriguing problems in cosmology. As a candidate for explaining the accelerated expansion of the universe, modified gravity theories have been intensively studied in recent years (see for reviews {\it e.g.} \cite{Clifton:2011aa,Tsujikawa:2011aa}). 
The most simplest way to modify Einstein's gravity could be to introduce an additional scalar degree of freedom in general relativity, called scalar-tensor theories of gravity. Although a number of the scalar-tensor theories have been so far proposed in various contexts, an interesting scalar-tensor theory would be the galileon theory \cite{Nicolis:2008aa}, whose Lagrangian contains higher, that is at least second, derivatives of the scalar field. Surprisingly, its special structure prevents third or higher derivative terms from appearing in the equation of motion for a scalar field, which in general leads to a ghost mode, hence it is free of Ostrogradski instability.
The galileon field have been discovered in the context of an effective field theory of modified gravity theories. For example, the cubic galileon field is originated from a brane bending mode in the decoupling limit \cite{Nicolis:aa,2003JHEP...09..029L} of the Dvali-Gabadadze-Poratti brane-world scenario \cite{2000PhLB..485..208D}, 
and the quartic and quintic galileon interactions as well as the cubic galileon are subsequently found  in the decoupling limit of the de Rham-Gabadadze-Tolley massive gravity \cite{2010PhRvD..82d4020D,Rham:2011aa}. The galileon self-interactions can be generalized in a curved space-time requiring that equations of motion for both scalar field and gravity remain at most second-order differential equations \cite{2009PhRvD..79h4003D}, and the most general scalar-tensor theory, whose equations of motion contains up to the second derivatives of the fields, is called the Horndeski theory \cite{Horndeski:1974aa,Deffayet:2011aa,2011PThPh.126..511K}. However, it was pointed out that one can further generalize these constructions without a ghost mode by allowing higher order derivatives in the equations of motion, dubbed as the beyond Horndeski or the GLPV theory \cite{2015PhRvL.114u1101G}. Furthermore, the recent works in \cite{2016JCAP...02..034L} showed that more general constructions of scalar-tensor theories is possible as long as they satisfy a degeneracy condition to eliminate the would-be Ostrogradski mode. Such degenerate theories have been also investigated based on Hamiltonian formulation in \cite{2015arXiv151206820L,2016JCAP...04..044C},  confirming the presence of an additional primary constraint which is necessary to eliminate the Ostrogradski mode. The detailed classification and the cosmological applications of each extended scalar-tensor theory have been investigated in \cite{2016arXiv160208398B} and \cite{2016arXiv160408638D}, respectively.

Another attempt to modify general relativity is to introduce an Abelian vector field. In massive gravity theory, a massive graviton carries five degrees of freedom, and one can separate out the tensor part and the vector part by introducing the St\"uckelberg field, $h_{\mu\nu} \to h_{\mu\nu}+\partial_\mu A_\nu +\partial_\nu A_\mu$. The vector Lagrangian in the decoupling limit of massive gravity does not only contain the standard Maxwell Lagrangian but also includes fully nonlinear corrections \cite{2013PhRvD..88h4003G,2013JHEP...11..059O}.
Motivated by this fact, vector-tensor theory is still a natural extension as an effective theory of gravity. 
In the pioneer work \cite{2014JHEP...04..082D}, the authors showed no-go theorem which states that the galileon-like terms are not allowed in a flat space-time under the assumption of  $U(1)$ invariance, while in a curved space-time the vector Horndeski term \cite{1976JMP....17.1980H} is allowed. However, once we relax the condition of $U(1)$ invariance, the vector field is no longer massless, and a finite number of interactions can be found. This extended theory of the massive vector field is called the generalized Proca (GP) theory \cite{2014JCAP...05..015H} which consists of six interaction terms in a flat space-time. The higher order Lagrangian can be similarly constructed in the same manner \cite{2016JCAP...02..004A}, however it identically vanishes by virtue of the Newton's identities as in the flat galileon \cite{2016PhLB..757..405B}. Covariantization procedure of the GP theory in a flat space-time can be similarly done as in the case of the generalized galileon. Furthermore, one can construct "beyond" type vector derivative interactions (beyond GP theory) as in the beyond Horndeski theory \cite{2016arXiv160505565H}. Summarizing these results, the vector-tensor theory investigated so far is described by the following action,
\ba
S_{\rm VT} = \int d^4x \, \sqrt{ -g } \,
\[\,
\sum_{n=2}^6  \Lag_n +\sum_{m=4}^6 \Lag_m^{\rm (B)}\],
\ea
where the GP interactions $\Lag_n$ are given by
\begin{subequations}
\ba
 \Lag_2&=& G_2(Y, F^2,  F\tilde{F}, (AF)^2 ),\\
\Lag_3&=& G_3(Y) \nabla_\mu A^\mu,\\
\Lag_4&=& G_4(Y) R -2 G_{4,Y}[(\nabla_\mu A^\mu)^2-\nabla_\mu A_\nu \nabla^\nu A^\mu],\\
\Lag_5&=& G_5(Y) G_{\mu\nu} \nabla^\mu A^\nu +{1\over 3} G_{5,Y}
[(\nabla_\mu A^\mu)^3-3\nabla_\mu A^\mu \nabla_\rho A_\sigma \nabla^\sigma A^\rho +2 \nabla_\rho A_\sigma \nabla^\gamma A^\rho \nabla^\sigma A_\gamma]~~~~~\\
&&-{\tilde G}_5(Y) {\tilde F}^{\alpha\mu}{\tilde F}^\beta_{~\mu}\nabla_\alpha A_\beta,\label{L5GP}\\
\Lag_6&=& G_6(Y) L^{\mu\nu\alpha\beta} \nabla_\mu A_\nu \nabla_\alpha A_\beta 
- G_{6,Y} {\tilde F}^{\alpha\beta}{\tilde F}^{\mu\nu} \nabla_\alpha A_\mu \nabla_\beta A_\nu,\label{L6GP}
\ea
\end{subequations}
and the beyond GP interactions $\Lag_m^{\rm (B)}$ are given by 
\begin{subequations}
\ba
\Lag_4^{\rm (B)}&=& G_4^{\rm (B)} (Y)\, \eps^{\mu\nu\rho\sigma}\eps^{\alpha\beta\gamma}_{~~~~\sigma} A_\mu A_\alpha \nabla_\nu A_\beta  \nabla_\rho A_\gamma,\label{L4beyond}\\
\Lag_5^{\rm (B)}&=& G_5^{\rm (B)} (Y)\, \eps^{\mu\nu\rho\sigma}\eps^{\alpha\beta\gamma\delta}A_\mu A_\alpha \nabla_\nu A_\beta  \nabla_\rho A_\gamma \nabla_\sigma A_\delta+{\tilde G}_5^{\rm (B)} \eps^{\mu\nu\rho\sigma}\eps^{\alpha\beta\gamma\delta}A_\mu A_\alpha \nabla_\nu A_\rho  \nabla_\beta A_\gamma \nabla_\sigma A_\delta,~~~~~~\\
\Lag_6^{\rm (B)}&=& G_6^{\rm (B)} (Y) \, \eps^{\mu\nu\rho\sigma}\eps^{\alpha\beta\gamma\delta} \nabla_\mu A_\nu \nabla_\alpha A_\beta  \nabla_\rho A_\gamma  \nabla_\sigma A_\delta.
\ea
\end{subequations}
Here, $G_{3,4,5,6}$, $G_{4,5,6}^{\rm (B)}$, ${\tilde G}_5$, and $\tilde{G}^{(B)}_5$
 are arbitrary functions of {\it the Proca mass term} $Y = A_\mu A^{\mu}$, $G_2$
 are arbitrary functions of $Y$, the Maxwell kinetic term ${F^2} = F_{\mu\nu}F^{\mu\nu}$, {$F\tilde{F} = F_{\mu\nu}\tilde{F}^{\mu\nu}$}, and ${(AF)^2}=A^\mu A^\nu F_\mu^{~\alpha} F_{\nu\alpha}$, where $F_{\mu\nu} = \nabla_{\mu} A_{\nu}-\nabla_{\nu} A_{\mu}$,  and $G_{i,Y}$ stands for $\partial G_i/ \partial Y$. 
The dual strength tensor and the double dual Riemann tensor are defined as ${\tilde F}^{\mu\nu} =\eps^{\mu\nu\alpha\beta}F_{\alpha\beta}/2$ and $L^{\mu\nu\alpha\beta}=\eps^{\mu\nu\rho\sigma} \eps^{\alpha\beta\gamma\sigma}R_{\rho\sigma\gamma\delta}/4$.
These Lagrangians with the replacement, $A_\mu \to \partial_\mu \phi$, coincide with the Horndeski and the beyond Horndeski Lagrangians which are invariant under a shift symmetry, $\phi \to \phi \,+\, {\rm constant}$.
This Horndeski structure guarantees that this $\phi$ field does not contain higher derivatives, meaning the absence of the Ostrogradsky ghost in the situation such that the vector field is simply replaced by the gradient of the $\phi$ field.
One should note that $U(1)$ gauge invariant Lagrangian can be found when the arbitrary functions $G_3$, $G_4$, $G_5$, and $G_6$ in the GP theory are set to be constant and $G_2=G_2({F^2, F{\tilde F}})$ and ${\tilde G}_5=0$. The remaining Lagrangians, after integration by parts, are those associated with $G_2$, $G_4$, and $G_6$, and the massless vector field is described by the arbitrary function of the Maxwell kinetic term and $F{\tilde F}$, the Einstein-Hilbert term, and the vector Horndeski \cite{1976JMP....17.1980H}.
The cosmological applications of the GP and beyond GP theories have been recently studied in \cite{2016arXiv160505066D,2016JCAP...06..048D,2016arXiv160703175H}, and a Higgs mechanism and black hole solutions for these theories are discussed in \cite{2014arXiv1408.6871H,2016PhRvD..93f4012H} and \cite{2016CQGra..33q5007C}, respectively. 
 
Now we want to address a question\, : 
``Is this the most general vector-tensor theory without introducing an extra (would-be ghostly) degree of freedom ?\,". The answer is probably no. As can be seen in the case of  
scalar-tensor theories, one can further extend vector-tensor theories 
with a degeneracy condition in the kinetic matrix which can kill the would-be dangerous mode.
To this end, in the present paper we investigate the most general degenerate vector-tensor theories up to 
 quadratic order in the first derivatives of the vector field.

This paper is organized as follows. 
In section\,\ref{sec2}, we introduce vector-tensor theories whose Lagrangian contain up to two derivatives acting on the vector and metric tensor fields. Then, we derive a kinetic matrix and a degeneracy condition by the use of $3+1$ decomposition without gauge fixing. In section\,\ref{sec3}, we investigate the most general theories which satisfy the degeneracy condition and classify all the possible cases. We also link the GP and the beyond GP theories with our new theories utilizing conformal and disformal transformations as well as a vector field redefinition. 
Section\,\ref{sec4} is devoted to the conclusion.
Details on metric and vector field transformations are summarized in appendix\,\ref{appendix-trf.}.
The explicit expressions of determinant of kinetic matrices in the cases A and B
 are collected in  appendix\,\ref{appendix-detC}.
Finally, the analysis of the case C, 
where the Einstein-Hilbert term is absent,
is performed in appendix\,\ref{appendix-CaseC}.

\section{Vector-tensor theories}
\label{sec2}

\subsection{The action}
We consider a class of vector-tensor theories, whose generic action contains
up to two derivatives with respect to $g_{\mu\nu}$ and $A_\mu$\,:
\ba
\label{action}
S[g_{\mu\nu},A_\mu] \equiv \int d^4x \, \sqrt{ -g } \,\(f \, R + C^{\mu\nu\rho\sigma}\,  \nabla_\mu A_\nu \, \nabla_\rho A_\sigma 
+G_3 \nabla_{\mu}A^{\mu} + G_2
\),
\ea
where $R$ is the Ricci scalar and 
$\nabla_\mu$ represents a covariant derivative with respect to the space-time metric, $g_{\mu \nu}$.
The tensor $C^{\mu\nu\rho\sigma}$ depends on $g^{\mu\nu}$, $A^\mu$ and $\eps^{\mu\nu\rho\sigma}$, which is defined as
\footnote{The symmetrization and anti-symmetrization are normalized by
\ba
	T^{(\mu\nu)} = \frac{1}{2}(T^{\mu\nu}+T^{\nu\mu}) \,, \qquad
	T^{[\mu\nu]} = \frac{1}{2}(T^{\mu\nu}-T^{\nu\mu}) \,. \notag
\ea
}
\ba
	C^{\mu\nu\rho\sigma} &=&
	\alpha_1 g^{\mu(\rho}g^{\sigma)\nu} + \alpha_2 g^{\mu\nu}g^{\rho\sigma}
	 + \frac{1}{2}\alpha_3 (A^\mu A^\nu g^{\rho\sigma} + A^\rho A^\sigma g^{\mu\nu}) \notag\\
	&& \quad
	  + \frac{1}{2} \alpha_4 (A^\mu A^{(\rho} g^{\sigma)\nu} + A^\nu A^{(\rho} g^{\sigma)\mu}) 
	+ \alpha_5 A^{\mu} A^{\nu} A^{\rho} A^{\sigma}+ \alpha_6 g^{\mu[\rho}g^{\sigma]\nu}\notag\\
	&& \quad
	+ \frac{1}{2}\alpha_7 (A^\mu A^{[\rho} g^{\sigma]\nu} - A^\nu A^{[\rho} g^{\sigma]\mu})
	+ \frac{1}{4}\alpha_8 (A^\mu A^\rho g^{\nu\sigma} - A^{\nu}A^{\sigma} g^{\mu\rho})
	+{1\over 2}\alpha_9 \eps^{\mu\nu\rho\sigma}.
	\label{Ctensor}
\ea
Here, we introduced arbitrary functions of $Y = A_\mu A^\mu$
 such as $f=f(Y)$, $\alpha_i=\alpha_i(Y)$, $G_3= G_3(Y)$ and $G_2= G_2(Y)$.
With our best knowledge, it will be the first time in the literature that the term with $\alpha_8$ is taken into account,
  which induces the cross term of $F_{\mu \nu}$ and $S_{\mu \nu}$ as seen in (\ref{LagSF}). We note that another possible term like $A^{\mu}A^{\nu}\nabla_{\mu}A_{\nu}$ can be absorbed into $G_3$ after integration by parts\,:   
   \ba
    {\cal G}_{3, Y} (Y) A^{\mu}A^{\nu} \nabla_{\mu}A_{\nu}
     = - \frac{1}{2} {\cal G}_3 (Y) \nabla_{\mu}A^{\mu}
      + \frac{1}{2}\nabla_{\mu} \bigl( {\cal G}_{3} (Y) A^{\mu} \bigr)\,,
   \label{red-yoshida}
   \ea
   where the subscript $, Y$ represents a derivative with respect to $Y$. 
   Furthermore, one might notice that the non-minimal coupling of the vector field to gravity, ${\cal G}_4(Y) G_{\mu\nu} A^\mu A^\nu$, should be systematically introduced since it carries two derivatives. However,  this term can be similarly absorbed into $C^{\mu\nu\rho\sigma}$ term up to a total derivative, thanks to the identity, $(\nabla_{\mu}\nabla_{\nu}-\nabla_{\nu}\nabla_{\mu}) A_{\rho} = R_{\mu\nu\rho}{}^{\sigma}A_\sigma$.
        
For later convenience, let us rewrite the vector self-interactions in terms of the symmetric and anti-symmetric parts of $\nabla_\mu A_\nu$, 
\ba
	S_{\mu\nu} = \nabla_{\mu} A_{\nu}+\nabla_{\nu} A_{\mu}, 
	\qquad 
	F_{\mu\nu} = \nabla_{\mu} A_{\nu}-\nabla_{\nu} A_{\mu}, 
\ea
and we then have 
\ba
4 \,C^{\mu\nu\rho\sigma} \nabla_{\mu}A_{\nu} \nabla_{\rho}A_{\sigma}
&=& 
\alpha_1 S_{\mu\nu}S^{\mu\nu} + \alpha_2 (S_\mu{}^{\mu})^2 \notag
+ \alpha_3 A^\mu A^{\nu} S_{\mu\nu}S_{\rho}{}^{\rho} \notag\\
&& \quad
+ \alpha_4 A^{\mu}A^{\nu}S_{\mu\rho}S_{\nu}{}^{\rho} 
+ \alpha_5 (A^{\mu}A^{\nu}S_{\mu\nu})^2+ \alpha_6 F_{\mu\nu}F^{\mu\nu} \notag\\
&& \quad
+ \alpha_7 A^\mu A^{\nu} F_{\mu\rho}F_{\nu}{}^{\rho}
+ \alpha_8 A^{\mu} A^{\nu}F_{\mu}{}^{\rho}S_{\nu\rho}
+\alpha_9 F_{\mu\nu} {\tilde F}^{\mu\nu}\,.
\label{LagSF}
\ea
One might think that it is possible to add terms related to the dual of $F_{\mu\nu}$ such as 
$A^\mu A^{\nu} F_{\mu\rho}{\tilde F}_{\nu}{}^{\rho}$ and $A^{\mu} A^{\nu}{\tilde F}_{\mu}{}^{\rho}S_{\nu\rho}$. 
The first term can be rewritten as
\ba
A^\mu A^{\nu} F_{\mu\rho}{\tilde F}_{\nu}{}^{\rho} 
= {1\over 4} Y F_{\mu\nu} {\tilde F}^{\mu\nu},
\label{otherterm}
\ea
where we have used the identity
 $F^{\mu\rho}{\tilde F_{\nu\rho}} = (F_{\alpha\beta} {\tilde F}^{\alpha\beta}/4) \, \delta^\mu_{~\nu}$, 
derived in \cite{Fleury:2014qfa,2016arXiv160508355A}. 
Thus, this term is already included in (\ref{LagSF}). 
Furthermore, it is easy to show that the second term $A^{\mu} A^{\nu}{\tilde F}_{\mu}{}^{\rho}S_{\nu\rho}$ 
 can be rewritten as 
\begin{align}
 {\cal G}_{9, Y} (Y) 
A^{\mu} A^{\nu}{\tilde F}_{\mu}{}^{\rho}S_{\nu\rho}
 &= \frac{1}{2} {\cal G}_9 (Y) F_{\mu\nu} {\tilde F}^{\mu\nu}
 + {\cal G}_{9, Y} (Y) A^\mu A^\nu F_{\mu \rho} {\tilde F}_{\nu}{}^{\rho}
 - \nabla_\mu \bigl( {\cal G}_9 (Y) A_\nu \tilde{F}^{\mu \nu} \bigr) \notag\\
 &= \left( \frac{1}{2} {\cal G}_9(Y) + \frac{1}{4} Y {\cal G}_{9, Y}(Y) \right)  F_{\mu \nu} {\tilde F}^{\mu \nu}
 - \nabla_\mu \bigl( {\cal G}_9 (Y) A_\nu \tilde{F}^{\mu \nu} \bigr) 
\,.
\label{IntbyParts}
\end{align}
 where we have used (\ref{otherterm}) in the second equality. Therefore, all the possible terms related to the dual of $F_{\mu\nu}$ can be totally absorbed into $\alpha_9 F_{\mu\nu} {\tilde F}^{\mu\nu}$.
In light of these facts, $C^{\mu\nu\rho\sigma}$ in (\ref{Ctensor}) is the most general form, which is composed by $A^\mu$, $g^{\mu\nu}$, and $\varepsilon^{\mu\nu\rho\sigma}$.

In the GP and the beyond GP theories, the $\alpha_6$ and $\alpha_7$ terms in (\ref{Ctensor}) was classified to $\Lag_2$ as an analogue of 
the k-essence term \cite{2000PhRvD..62b3511C,2000PhRvL..85.4438A} in scalar field theories
because these terms do not carry the dangerous time-derivative of the time-component, $A_0$.
However, in more general set-up, these arbitrary functions are also responsible for satisfying a degeneracy condition, and hence those contributions in addition to the one from $\alpha_8$ must be properly taken into account as we will discuss this issue in the next section. 

Before the end of this subsection, we discuss the connection with scalar-tensor theories. As discussed in \cite{2014JCAP...05..015H}, a subclass of the GP theory can be obtained from the shift-symmetric Horndeski theory, which enjoys a symmetry of the action under $\phi \to \phi + const.$\,, just by replacing $\nabla_\mu \phi$ with $A_\mu$.
However, our theory of the vector field {\it cannot} be obtained from
 any kinds of scalar field theories (even the ones recently formulated in  \cite{2016JCAP...04..044C,2015arXiv151206820L}) via such direct replacement of the fields.
Since the indices of $\nabla_\mu \nabla_\nu \phi$ is symmetric, only the symmetric parts of $C^{\mu\nu\rho\sigma}$ in (\ref{action}) and (\ref{Ctensor}) can survive in the case of scalar field\,:
\ba
	C^{\mu\nu\rho\sigma} (\nabla_\mu \nabla_\nu \phi) (\nabla_\rho \nabla_\sigma \phi)
	&=& C^{(\mu\,\nu) \, (\rho\,\sigma)} (\nabla_\mu \nabla_\nu \phi) (\nabla_\rho \nabla_\sigma \phi).
\ea
As a matter of fact, even in the case of the GP theory, ${\tilde G}_5$ term in (\ref{L5GP}) and $G_6$ term in (\ref{L6GP}) cannot be also obtained from the shift-symmetric Horndeski theory since they are identically zero \cite{2014JCAP...05..015H,2016JCAP...02..004A,2016PhLB..757..405B}.
To put it another way, since the anti-symmetric part of $C^{\mu\nu\rho\sigma}$ is automatically dropped off
 in the theory of the scalar field, we cannot explore a whole class of vector field theories where those anti-symmetric parts can play an important role. 

\subsection{3+1 Decomposition}
In order to write down a degeneracy condition for the kinetic matrix, 
we need to extract the time derivative component of the action (\ref{action}). 
A convenient way to do this without gauge fixing is $3+1$ decomposition. To the end, 
we assume space-time manifold is foliated by spacelike hypersurfaces $\Sigma_{t}$.
Let us define normal vector $n^{\mu}$ of each hypersurface $\Sigma_t$, which satisfies $n^{\mu}n_{\mu} = -1$. The induced metric on $\Sigma_t$, $\gamma_{\mu\nu}$, are defined by
\begin{align}
\gamma_{\mu\nu} = g_{\mu\nu} + n_{\mu} n_{\nu}.
\end{align}
Any covariant tensor fields can be decomposed by this induced metric 
and the normal vector.
For example, $A_{\mu}$ can be decomposed as
\begin{align}
A_{\mu} &= - n_{\mu} A_{*}  + \hat{A}_{\mu},
\end{align}
where we define
\begin{subequations}
\begin{align}
A_{*} &:= n^{\mu}A_{\mu},\\
\hat{A}_{\mu} &:= \gamma_{\mu}{}^{\nu}A_{\nu}.
\end{align}
\end{subequations}
The derivative of the normal vector can also be decomposed into the extrinsic curvature $K_{\mu\nu}$ and acceleration vector $a_\mu$,
\begin{align}
\nabla_{\mu} n_{\nu} = - n_{\mu} a_{\nu} + K_{\mu\nu} ,
\end{align}
where the extrinsic curvature and the acceleration vector are defined by
\begin{subequations}
\begin{align}
a^{\mu} &:= n^{\nu}\nabla_{\nu}n^{\mu},\\
K_{\mu\nu} &:= \gamma_{\mu}{}^{\rho} \gamma_{\nu}{}^{\sigma}\nabla_{\rho}n_{\sigma}.
\end{align}
\end{subequations}
Then, the first derivative of the vector field reads
\ba
\nabla_{\mu}A_{\nu} &=& n_{\mu} n_{\nu} (\dot{A}_{*}-a^{\rho}\hat{A}_{\rho} )  
+ n_{\mu} (- \dot{\hat{A}}_{\nu} + K_{\nu}{}^{\rho} \hat{A}_{\rho} + a_{\nu}A_{*}  ) \notag\\
 &&
 +(K_{\mu}{}^{\rho}\hat{A}_{\rho} - D_{\mu}A_{*})n_{\nu} 
 +D_{\mu}\hat{A}_{\nu}  -K_{\mu\nu}A_{*},
\ea
where
$D_\mu$ represents a covariant derivative with respect to the spatial metric, $\gamma_{\mu \nu}$ and
a dot represents the Lie derivative along $n^{\mu}$:
\begin{subequations}
\begin{align}
\dot{A}_{*} &= \mathsterling_{n} A_{*} = n^{\mu} \nabla_{\mu} A_{*},\\
\dot{\hat{A}}_{\mu} &= \mathsterling_{n} \hat{A}_{\mu} = n^{\nu}\nabla_{\nu} \hat{A}_{\mu} + \hat{A}_{\nu}\nabla_{\mu}n^{\nu} \,.
\end{align}
\end{subequations}
The kinetic part of the Lagrangian (\ref{action}) can be expressed in terms of 
$A_{*} $, $\hat{A}_{\mu}$, $\gamma_{\mu\nu}$, and $K_{\mu\nu}$
as  
\ba
\Lag_{\rm kin}={\cal A} \Asd^2 +2\B^i \Asd \Ahd_\mu +2\C^{\mu\nu} \Asd K_{\mu\nu}+ \D^{\mu\nu} \Ahd_\mu \Ahd_\nu
+2\E^{\mu\nu\rho} \Ahd_\mu K_{\nu\rho} + \F^{\mu\nu\rho\sigma} K_{\mu\nu} K_{\rho\sigma},
\ea
with
\begin{subequations}
\begin{align}
{\cal A} =&  \alpha_1 + \alpha_2 -  (\alpha_3+\alpha_4) A_*^2+\alpha_5 A_{*}^4,\\
{\cal B}^{\mu} =& -\frac{1}{4}\hat{A}^{\mu}A_{*}(-2 \alpha_3 -2 \alpha_4 +\alpha_8 + 4\alpha_5 A_{*}^2),\\
{\cal C}^{\mu\nu} =&  {1\over 2}A_{*}(-\alpha_3 -2 \alpha_4 +2 \alpha_5 A_{*}^2 )\hat{A}^{\mu}\hat{A}^{\nu} +{1\over 2} A_{*}(4f_Y+2\alpha_2 -\alpha_3 A_{*}^2)\gamma^{\mu\nu},\\
{\cal D}^{\mu\nu} =& -\frac{1}{4}(\alpha_4 + \alpha_7-\alpha_8-4\alpha_5 A_{*}^2) \hat{A}^{\mu}\hat{A}^{\nu}+\frac{1}{4}(-2\alpha_1-2\alpha_6+A_{*}^2(\alpha_4+\alpha_7+\alpha_8))\gamma^{\mu\nu},\\
{\cal E}^{\mu\nu\rho} =&  \alpha_1 \gamma^{\mu(\nu}\hat{A}^{\rho)}+{1\over 2}(-4f_Y+\alpha_3 A_{*}^2) \hat{A}^{\mu}\gamma^{\nu\rho} + \frac{1}{4}\hat{A}^{\mu}\hat{A}^{\nu}\hat{A}^{\rho} (2\alpha_4 - \alpha_8 -4 \alpha_5 A_{*}^2),\\
{\cal F}^{\mu\nu\rho\sigma}= &
(f + \alpha_1 A_{*}^2)  \gamma^{\mu(\rho}\gamma^{\sigma)\nu} + (-f+\alpha_2 A_{*}^2)  \gamma^{\mu\nu}\gamma^{\rho\sigma} \notag + {1\over 2} (4f_Y- \alpha_3 A_{*}^2) (\hat{A}^\mu \hat{A}^\nu \gamma^{\rho\sigma} + \hat{A}^\rho \hat{A}^\sigma \gamma^{\mu\nu}) \\
& - \alpha_1 (\hat{A}^\mu \hat{A}^{(\rho} \gamma^{\sigma)\nu} + \hat{A}^\nu \hat{A}^{(\rho} \gamma^{\sigma)\mu} )+ 2 (-\alpha_4 +  \alpha_5 A_{*}^2) \hat{A}^{\mu} \hat{A}^{\nu} \hat{A}^{\rho} \hat{A}^{\sigma},
\end{align}
\end{subequations}
 where we have also used
\ba
R &=& {}^{(3)}R + K_{\mu\nu}K^{\mu\nu}-K^2 -2 \nabla_{\mu}(a^\mu -K n^{\mu} ) \,,
\ea
and ${}^{(3)}R$ stands for the three-dimensional Ricci scalar composed by the spatial metric, $\gamma_{\mu \nu}$, 
and $K=\gamma^{\mu\nu}K_{\mu\nu}$.
Note that the kinetic Lagrangian does not depend on $\alpha_9$. 
This is because 
the term $\alpha_9 F_{\mu \nu} \tilde{F}^{\mu \nu}$ only contains up to the first Lie derivative of $\Ah_\mu$, {\it i.e.,} $\alpha_9$ can be freely chosen
in constructing degenerate theories.

\subsection{Block-diagonalized kinetic matrix}
The kinetic Lagrangian we obtained in the previous subsection is still involved to analyze the structure of vector-tensor theories. We would like to simplify this kinetic Lagrangian by changing the basis to the one, which can (even partially) diagonalize the kinetic matrix based on the irreducible representation. Following the pioneering work \cite{2015arXiv151206820L}, we introduce two unit spatial vectors $u_{\mu}$ and $v_{\mu}$, which satisfy
\ba
u^\mu u_\mu = v^\mu v_\mu = 1, \quad u^\mu v_\mu = v^\mu \Ah_\mu = \Ah^\mu u_\mu = 0 \, ,
\ea
and
\begin{align}
 n^{\mu}u_{\mu} = n^{\mu} v_{\mu} = 0.
\end{align}
The vector $\Ah_\mu$ can be normalized by $\Ah_\mu/|\Ah|$
 where we define the inner product of the spatial component as $\Ah^2=\Ah_\mu \Ah^\mu$ 
and $|\Ah|$ represents $\sqrt{\Ah^2}$. 
Then, we can build orthogonal bases of the three dimensional vector space on $\Sigma_t$,
$V^a_\mu$ ($a$ runs from $1$ to $3$) as 
\ba
V_\mu^1= {\Ah_\mu \over |\Ah|}, \quad ~ V_\mu^2= u_\mu, \quad ~ V_\mu^3= v_\mu,
\ea
which satisfy $V^{a}_{\mu}V^{b\mu} = \delta^{ab}$ and 
$\gamma^{\mu}{}_{\nu} = \delta_{ab} V^a{}^{\mu}V^b{}_{\nu}$.
By using these unit vectors, one can construct the following $6$ independent symmetric matrices $U^I_{\mu\nu}$
($I$ runs from $1$ to $6$), 
\ba
U^1_{\mu\nu} = \frac{1}{\Ah^2} \Ah_\mu \Ah_\nu, \quad 
U^2_{\mu\nu} &=& \frac{1}{\sqrt{2}} (\gamma_{\mu\nu} - U^1_{\mu\nu}), \quad \quad
U^3_{\mu\nu} = \frac{1}{\sqrt{2}}(u_\mu u_\nu - v_\mu v_\nu), \notag\\
U^4_{\mu\nu} = \frac{1}{\sqrt{2}}(u_\mu v_\nu + u_\nu v_\mu) ,  \quad 
U^5_{\mu\nu} &=& \frac{1}{ \sqrt{2}\, |\Ah| }(u_\mu \Ah_\nu + u_\nu \Ah_\mu), \quad 
U^6_{\mu\nu} = \frac{1}{\sqrt{2 }\,|\Ah| }(v_\mu \Ah_\nu + v_\nu \Ah_\mu), \quad
\ea
{which satisfy $U^{I}_{\mu\nu}U^{J}{}^{\mu\nu} = \delta^{IJ}$}.
{Since $U^{I}_{\mu\nu}$ span the space of symmetric tensors on $ \Sigma_t $, these tensors satisfy the relations $\gamma^{\mu}{}_{(\rho}\gamma^{\nu}{}_{\sigma)} = \delta_{IJ} U^{I}{}^{\mu\nu} U^{J}{}_{\rho\sigma}$}.
Then, we can decompose $\Ahd_\mu$ and $K_{\mu\nu}$
 along these vector and tensor bases\,:  
 \ba
 \Ahd_\mu = {\dot A}_a \, V_\mu^a,
 \quad ~ 
K_{\mu\nu} = K_I \, U_{\mu\nu}^I.
\ea
Each coefficient such as $\dot{A}_a$ and $K_I$ can be obtained
 by projecting $\Ahd_\mu$ and $K_{\mu \nu}$ onto those bases,
$\dot{A}_a = V^a{}^{\mu} \dot{\hat{A}}_\mu$ and $K_I = U^I{}^{\mu\nu} K_{\mu\nu}$.
In terms of $V_\mu^a$ and $U_{\mu \nu}^I$, one can construct the scalar, vector, and tensor quantities 
 which transform as scalar, vector, and tensor respectively under a rotation around the axis, $\hat{A}_{\mu}$.
 For example, the contraction of any vector field with $V_\mu^1$ extract the scalar component in the vector field
  while that with $V_\mu^{2, 3}$ yields the vector components.
  
Now the kinetic Lagrangian is rewritten as
\ba
\Lag_{\rm kin}&=&{\cal A} \Asd^2 +2\B \Asd \Ad_1 
+2\Asd  (\C_1 K_1+\C_2 K_2)
+ \D_1 \Ad_1 ^2 + \D_2 (\Ad_2 ^2 +\Ad_3 ^2 )\notag\\
&&
+2\Ad_1(\E_1 K_1 +\E_2  K_2)+2\E_3 (\Ad_2 K_5+\Ad_3 K_6 )\notag\\
&&
+ \F_1K_1^2 +\F_2 K_2^2 +2\F_3 K_1K_2 +\F_4(K_3^2+K_4^2)+\F_5(K_5^2+K_6^2),
\ea
where the coefficients are given by 
\begin{subequations}
\begin{align}
\A&=  \alpha_1 + \alpha_2 -  (\alpha_3+\alpha_4) A_*^2+\alpha_5 A_{*}^4,\\
\B&= {1\over 4} (2\alpha_3+2\alpha_4-\alpha_8-4\alpha_5 \As) \As |\Ah|,\\
\C_1&= {1\over 2}  \(4f_Y + 2\alpha_2-\alpha_3 \As^2 -(\alpha_3+2\alpha_4)\Ah^2+2\alpha_5 \As^2 \Ah^2\)\As,\\
\C_2&= {1\over \sqrt{2}}(4f_Y+2\alpha_2-\alpha_3\As^2) \As, \displaybreak[1]\\
\D_1&= -{1\over 4} \( 2(\alpha_1+\alpha_6)+(\alpha_4+\alpha_7-\alpha_8)\Ah^2-(\alpha_4+\alpha_7+\alpha_8)\As^2-4\alpha_5\As^2\Ah^2\),\\
\D_2&= -{1\over 4}  \( 2(\alpha_1+\alpha_6)-(\alpha_4+\alpha_7+\alpha_8)\As^2\),
\displaybreak[1]
\\
\E_1&= -{1\over 4} |\Ah| \(8f_Y -4\alpha_1-(2\alpha_4-\alpha_8)\Ah^2-2\alpha_3\As^2+4\alpha_5\As^2\Ah^2\),\\
\E_2&= -{1\over \sqrt{2}} (4f_Y-\alpha_3 \As^2) |\Ah|,\\
\E_3&={1\over \sqrt{2}}\alpha_1 |\Ah|,  \\
\F_1&= (\alpha_1+\alpha_2)\As^2+(4f_Y-2\alpha_1)\Ah^2-\alpha_3\As^2\Ah^2-\alpha_4\Ah^4+\alpha_5\As^2\Ah^4,\\
\F_2&= -f+(\alpha_1+2\alpha_2)\As^2,\\
\F_3&= -{1\over \sqrt{2}}\(2f-4f_Y \Ah^2-2\alpha_2 \As^2+\alpha_3\As^2\Ah^2\),\\
\F_4&=f+\alpha_1 \As^2,\\
\F_5&=f+\alpha_1\As^2-\alpha_1\Ah^2.
\end{align}
\end{subequations}
We can further rewrite this kinetic Lagrangian by using two $4 \times 4$ matrices and $2 \times 2$ matrix,
\ba
\Lag_{\rm kin}=
\left(
\begin{array}{ccc}
	{\bm m}^T & {\bm m}_1^T & {\bm m}_2^T \\
\end{array}
\right)
\left(
\begin{array}{ccc}
	{\cal M} & 0 & 0 \\
	0 & {\cal M}_1 & 0 \\
	0 & 0 & {\cal M}_2 \\
\end{array}
\right)
\left(
\begin{array}{ccc}
	{\bm m} \\
	{\bm m}_1  \\
	{\bm m}_2  \\
\end{array}
\right)
,
\label{Lkin}
\ea
where we defined the component column matrices ${\bm m}=\{\Asd, \Ad_1, K_1, K_2\}$, 
${\bm m}_1=\{\Ad_2, \Ad_3, K_5, K_6\}$, and ${\bm m}_2=\{K_3, K_4\}$,
 and a matrix with a superscript $T$ represents the transposed matrix.
The matrices $\M$, $\M_1$, and $\M_2$ describe the kinetic matrices of the scalar, vector and tensor sectors, respectively  
whose explicit forms are given by
\ba
{\cal M}=\left(
\begin{array}{cccc}
\A & \B & \C_1  & \C_2\\
\B & \D_1 & \E_1 & \E_2 \\
\C_1 & \E_1 & \F_1& \F_3\\
\C_2 &\E_2 & \F_3 &\F_2
\end{array}
\right), \quad 
{\cal M}_1=\left(
\begin{array}{cccc}
	\D_2 & 0 & \E_3  & 0\\
	0 & \D_2 & 0 & \E_3 \\
	\E_3 & 0 & \F_5& 0\\
	0 &\E_3 & 0 &\F_5
\end{array}
\right),\quad 
{\cal M}_2=\left(
\begin{array}{cc}
	\F_4 & 0 \\
	0 & \F_4 \\
\end{array}
\right).
\label{kineticM}
\ea
Interestingly, the scalar, vector, and tensor sectors are never mixed due to the different transformation property, and they can be treated independently as can be seen in (\ref{Lkin}).
Note that this is true only if the Lagrangians contains up to two derivatives with respect to space-time, {\it i.e.,} the kinetic part of the Lagrangians is strictly quadratic.

In the construction of degenerate vector-tensor theories, we need to remove the extra degree of freedom in the scalar sector,
which typically contains 
the Ostrogradsky ghost.
This is exactly the same situation as in the scalar-tensor theories, however the crucial difference in here is the presence of the $\Ad_1$ components in the kinetic matrix $\M$. Due to this additional component in the kinetic matrix, the degeneracy condition in the scalar-tensor theories, which removes the ghostly degree of freedom, will no longer satisfy the degeneracy condition in the vector-tensor theories in general.

\subsection{Metric and vector field transformations}
Metric and vector field transformations are potent tools to investigate the basic properties of gravitational theories.
We here consider new types of metric transformations by invoking a vector field\footnote{A disformal transformation of the vector field in a Minkowski background, $\eta_{\mu\nu} \to \eta_{\mu\nu} + \Gamma(Y) A_\mu A_\nu$, has first introduced in \cite{2016PhLB..757..405B}.}.
As a natural extension of the metric transformation with a scalar field,
the new transformation is given by
\begin{align}
 g_{\mu\nu} \quad \rightarrow \quad
 \bar{g}_{\mu\nu} = \Omega(Y) \, g_{\mu\nu} + \Gamma(Y) \, A_{\mu}A_{\nu}.
 \label{gtilde}
\end{align}
 Here, we introduced  
 the conformal factor $\Omega$ and the disformal factor $\Gamma$ which are
  functions of $Y$ while they can be functions of $\phi$ and $X = g^{\mu \nu} \nabla_\mu \phi \nabla_\nu \phi$
  in the case of scalar field. Although it may be possible to include derivatives of the vector field in the disformal/conformal factors as in the case of the scalar field, 
we only focus on the most general derivative-independent metric transformation (\ref{gtilde}) in the present paper for simplicity. 
 
Moreover, let us introduce a field redefinition of the vector field by
\begin{align}
 A_{\mu} \quad \rightarrow \quad \bar{A}_{\mu} = \Upsilon(Y) \, A_{\mu},
 \label{Atilde}
\end{align}
 where $\Upsilon$ is called a {\it rescaling factor}, which is a function of $Y$.
Apparently, there is no analog of this kind of the field redefinition in scalar field theories
since this is not a simple redefinition of $\phi$ nor $X$.
Throughout this paper, we assume that the metric transformations (\ref{gtilde}) and the vector field redefinition (\ref{Atilde}) satisfy $\Omega>0$ and $\Upsilon \neq 0$.  We also assume that the metrics in both frames have Lorentzian signatures, $\det {\bar g}_{\mu\nu}<0$ and $\det g_{\mu\nu}<0$, which translate into $\Omega + Y \Gamma > 0$ from (\ref{sqrtg}).

After straightforward but tedious computation, one can show that 
	the action (\ref{action}) is invariant under the metric transformation (\ref{gtilde}) and vector field redefinition (\ref{Atilde}) by redefining arbitrary functions such as $f$ and $\alpha_i$, that is, the form of the action reduces to the same form under these transformations.
The result is quite reasonable since the number of derivatives is preserved under these transformations. 
All the technical details associated with the metric and vector field transformations are summarized in appendix \ref{appendix-trf.}.

\section{Degenerate vector-tensor theories}
\label{sec3}
In this section, we derive a degeneracy condition of the kinetic matrix ${\cal M}$ in (\ref{kineticM}) to eliminate the dangerous mode, and then we focus on the classification of extended vector-tensor theories based on the degeneracy condition.

\subsection{Degeneracy condition}
We wish to find degenerate theories where the matrix $\M$ has at least one zero eigenvalue, 
which can be checked in the eigenvalue equation of $\M$, 
\ba
\det \, ({\cal M}-\lambda I)
= \det {\cal M} + y_1 \lambda + y_2 \lambda^2
+ y_3 \lambda^3 + \lambda^4 =0, 
\ea
where $I$ is the identity matrix and $\lambda$ is the eigenvalue. 
The necessary condition to remove the unwanted degree of freedom is 
given by $\det {\cal M}=0$, which implies the existence of a primary constraint (see for the detail in the context of classical mechanics \cite{2016JCAP...07..033M}). Then, the appropriate number of constraints should exist depending on whether this system is a first or second class, 
	and vector-tensor theories, which satisfies $\det {\cal M}=0$, have at most five degrees of freedom.
Furthermore, it is possible that $\M$ has two or more zero eigenvalues when $y_1$ (or subsequently $y_2$ and $y_3$) is zero.
One therefore needs to carefully check the number of zero eigenvalues 
to confirm the number of degrees of freedom.
As we will see, in order to have $y_1=0$, we additionally need to tune the arbitrary functions.
In the present paper, we mainly focus on the case in which only one of the eigenvalues is 
zero, {\it i.e.,} $\det \M=0$ but $y_{1,2,3} \neq 0$ as a generalization of the Proca theory.

We now want to write down $\det \M$ in the power series of $\As$. Since $f$ and $\alpha_i$ are functions of $Y$, we can re-express the matrix elements in terms of $Y$ and $\As$ by replacing $\Ah$ by its definition, $Y=-\As^2 + \Ah^2 $. 
The determinant of this matrix is formally given by the following form which contains up to $\As^4$, 
\ba
\det {\cal M} = D_0 (Y)+ D_1(Y) \As^2 + D_2(Y) \As^4.
\ea
Here, $D_0(Y)$ is given by 
\ba
D_0(Y)=\frac{1}{16} (\alpha _1+\alpha _2)\, 
Q(f, \alpha_1, \alpha_2, \alpha_4, \alpha_8, \beta),
\label{caseC}
\ea
 where
\ba
Q(f, \alpha_1, \alpha_2, \alpha_4, \alpha_8, \beta)
&=&8 f^2 \Bigl( 2 \alpha _1-\beta +(\alpha _4-\alpha _8)Y \Bigr)
+32 Y^2 f_Y^2 \Bigl( 2 \alpha _1-\beta +(\alpha_4+\alpha _8) Y \Bigr) \notag\\
&&
+  Y f \Bigl( 8 \alpha _1 \beta +16 f_Y (2 \alpha_1+\beta +\alpha _4 Y)
-192 f_Y^2+\alpha _8^2 Y^2+4\alpha _4\beta  Y \Bigr)\,, \qquad
\label{Qdef}
\ea
and we have introduced a convenient variable\,:
\ba
\beta=-2\alpha_6-\alpha_7 Y.
\ea
Since $\alpha_6$ and $\alpha_7$ 
always appear in this combination in the matrix ${\cal M}$,
the determinant is solely determined by $\beta$, not 
$\alpha_6$ and $\alpha_7$ independently.
On the other hand, ${\cal M}_1$ and ${\cal M}_2$ individually depend on both $\alpha_6$ and $\alpha_7$.

Then, $D_1$ is given by
\ba
D_1(Y)
&=&
{\cal W}_1 \, \alpha_8^2 +{\cal  W}_2\, \alpha_8 + {\cal W}_3,
\label{d1d0}
\ea
where  we defined
\begin{subequations}
\begin{align}
{\cal W}_1&=\frac{1}{8}(\alpha_1+\alpha_2)Y^2 f + \frac{1}{16} Y (f-\alpha _1 Y) \(2 f+(\alpha _1+3 \alpha _2) Y\),\\
{\cal W}_2&=\frac{1}{4}  (\alpha _1+\alpha _2)
 \(4(4 Y^2 f_Y^2 -f^2) + Y f \left(\alpha _1+\alpha _2+Y (\alpha _3+\alpha _4)+\alpha _5 Y^2\right)\) \notag\\
&~~~
-\frac{1}{16} Y (2 \alpha _2+4 f_Y+\alpha _3 Y) \(16 \alpha _1 Y f_Y+f (2 \alpha _2-12 f_Y+\alpha _3 Y)\), \displaybreak[1]\\
{\cal W}_3&=
\frac{1}{4} \left(\alpha _1+\alpha _2\right) \Bigl(\alpha _5 \bigl(\alpha _4 Y^3 f+Y f\left(-8 f+2 \alpha _1 Y-\beta  Y\right)\bigr)+\alpha _4^2 Y^2 f
\notag\\
& ~~~~~~~~~~~~~~~~~
+\alpha _4 Y \left(4 \beta  f+Y \left(-3 \alpha _1 \beta +4 \alpha _1 f_Y+16 f_Y^2\right)\right)
\notag\\
& ~~~~~~~~~~~~~~~~~
+2 \left(\alpha _1-2 f_Y\right) \left(Y \left(4 \left(\alpha _1+\beta \right) f_Y-3 \alpha _1 \beta \right)-2 f \left(\alpha _1-\beta -6 f_Y\right)\right)\Bigr)\notag\\
&~~~
+\frac{1}{2} \alpha _1 Y \left(\alpha _1 \beta  \left(2 \alpha _1+\alpha _4 Y\right)-2 f_Y \left(4 \alpha _1 \beta +4 f_Y \left(\alpha _4 Y-\beta \right)+\alpha _3 \alpha _4 Y^2-\alpha _3 \beta  Y+2 \alpha _1 \alpha _3 Y\right)\right)
\notag\\ &~~~
+\frac{1}{16}f \Bigl(8 f_Y \bigl(-12 \alpha _1^2+6 f_Y \left(2 \alpha _1-\beta +\alpha _4 Y\right)+2 \alpha _1 \left(3 \beta +\left(5 \alpha _3+\alpha _4\right) Y\right)+\alpha _3 Y \left(\alpha _4 Y-\beta \right)\bigr)\notag\\
&~~~~~~~~~~~~~~
+\beta  \left(-12 \alpha _1^2+\alpha _3^2 Y^2-4 \left(3 \alpha _3+4 \alpha _4\right) \alpha _1 Y\right)+\left(2 \alpha _1-\alpha _3 Y\right) \left(6 \alpha _1+\alpha _3 Y\right) \left(2 \alpha _1+\alpha _4 Y\right)\Bigr)\notag\\
&~~~
+\frac{1}{2} \left(\alpha _3+\alpha _4\right) f^2 \left(-2 \alpha _1+\beta +\alpha _3 Y\right).
\end{align}
\end{subequations}
Finally, while the expression of $D_2$ itself is quite involved,
	a linear combination of $D_0$, $D_1$, and $D_2$ takes a rather simple form as
	\ba
	D_1(Y)-Y D_2(Y)&=&-{1 \over 16Y} (f-\alpha_1 Y) \(2\alpha_1 + (\alpha_4+\alpha_8)Y-\beta\)\notag\\
	&&\times
	\[4 \(\alpha _1+\alpha _2+ (\alpha _3+\alpha _4)Y +\alpha _5 Y^2\) 
	\(2 f+(\alpha _1+3 \alpha _2) Y\)\notag\\
	&&~~~-3 Y (2 \alpha _2+4 f_Y+\alpha _3 Y){}^2 \] + {D_0(Y) \over Y}.
	\label{eq-DY3}
	\ea
Then, the degeneracy condition can translate into the following conditions, 
\ba
D_0(Y)=D_1(Y)=D_2(Y)=0.
\label{degeneracy}
\ea
In the subsection \ref{subsec:CaseA} and \ref{subsec:CaseB},
	we will use $D_1(Y)-Y D_2(Y)=0$ instead of $D_2(Y)=0$ as an independent condition in classifying degenerate vector-tensor theories since the expression is much simpler than $D_2(Y)$ itself.
We immediately notice that (\ref{caseC}) has two branches: $\alpha_1+\alpha_2=0$ and $Q(f, \alpha_1, \alpha_2, \alpha_4, \alpha_8, \beta)=0$ with $\alpha_1 + \alpha_2 \neq 0$. 
The former case corresponds to a broader class of the GP and the beyond GP theories since $\alpha_1+\alpha_2=0$ 
is also satisfied in both theories, while the latter case is a completely new class of vector-tensor theories.
In order to count the total degrees of freedom, we also need to check the degeneracy/non-degeneracy of the matrices $\M_1$ and $\M_2$
 to examine the presence of further additional primary constraint(s). 
For example, the determinant of $\M_2$ is zero only if $f=\alpha_1=0$
since it is solely determined by $\F_4=f+\alpha_1 \As^2$,
which corresponds to {\it no tensor modes},
and there is no gravitational wave in this case.
We also summarize the expression of $\M_1$ in appendix \ref{appendix-detC}. 

\begin{figure}[t]
	\begin{center}
		\includegraphics[width=160mm]{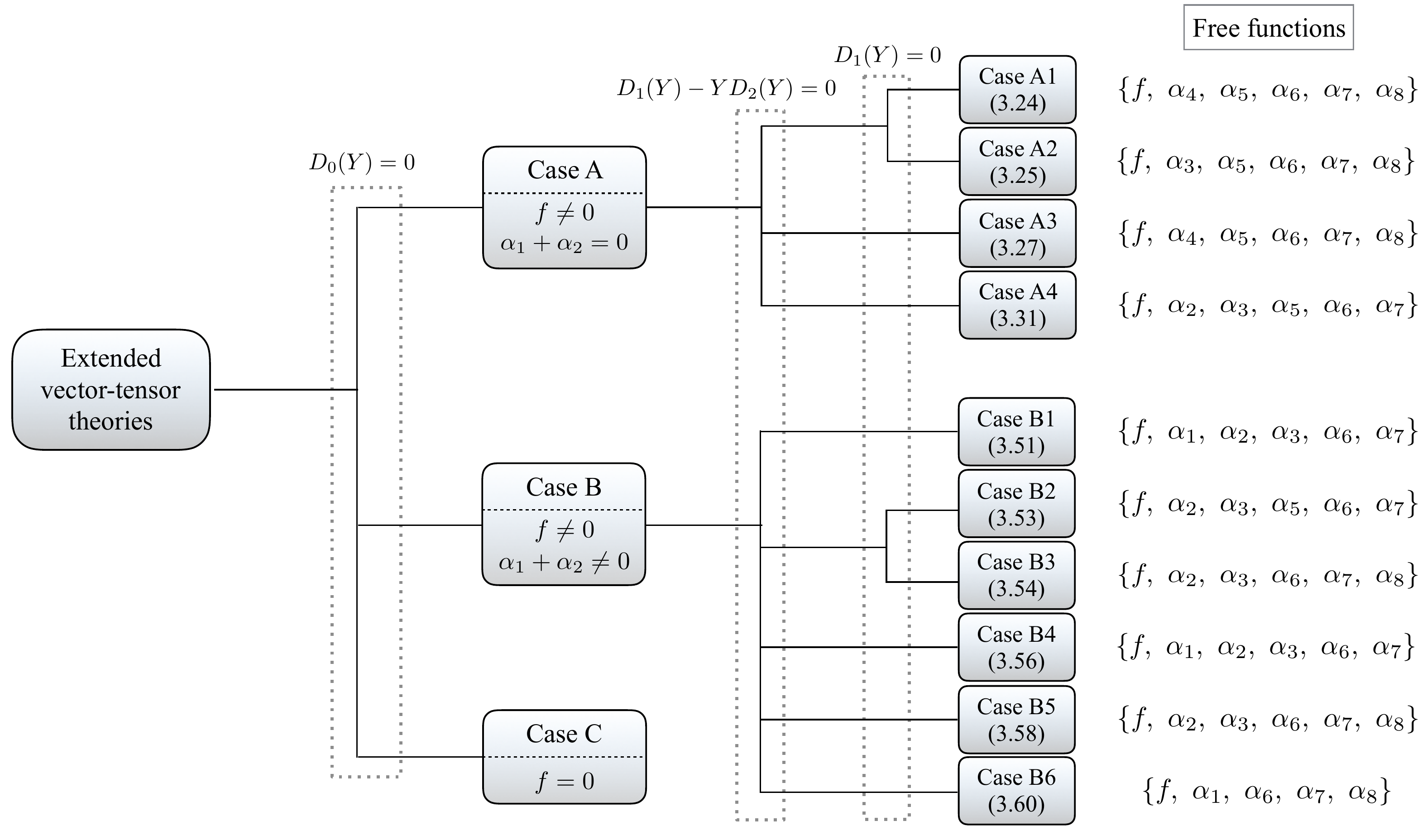}
	\end{center}
	\caption{
			Classification of extended vector-tensor theories, which satisfies the degeneracy condition (\ref{degeneracy}). 
			The theories with $f\neq 0$ is divided into the case A and the case B on the basis of the conditions, $\alpha_1+\alpha_2=0$ or $\alpha_1+\alpha_2\neq0$. 
			On the other hand, the theory with $f =0$, the case C, automatically satisfies the condition $D_0(Y)=0$. The further branches are obtained from the other conditions $D_1(Y)-Y D_2(Y)=0$ and $D_1(Y)=0$ in all cases. The detailed classification of the case C in appendix \ref{appendix-CaseC} is omitted, for simplicity.
			In each individual case, the example of the set of free functions is shown on the right side.
			Note that in the case A4 the free functions $f$ and $\alpha_{2,3,5}$ is required to satisfy the condition $W_2^2-4W_1W_3 \geq 0$.
	}
	\label{figure}
\end{figure}

The detailed analysis below proceeds as follows.
We first take a look at the special two cases, which correspond to the GP and the beyond GP theories in the next two subsections. 
Then, we investigate the general case and find all possible branches, which satisfy the degeneracy condition (\ref{degeneracy}).
In subsection \ref{subsec:CaseA}, we focus on the case $\alpha_1+\alpha_2=0$ denoted as the case A,
 and then the case $\alpha_1 + \alpha_2 \neq 0$ with
 $Q(f, \alpha_1, \alpha_2, \alpha_4, \alpha_8, \beta)=0$, denoted as the case B, will be investigated in the subsection \ref{subsec:CaseB}.
We here only consider $f \neq 0$ case, and the $f=0$ case are summarized in appendix \ref{appendix-CaseC}. 
We also summarize all results of $\det \M_1$ and $y_1$ for each case
in appendix \ref{appendix-detC}.
The schematic figure of the classification are shown in figure \ref{figure}. 

\subsection{Generalized Proca theory}
\label{sec:GP}
The action of the GP theory \cite{2014JCAP...05..015H}
up to the quartic Lagrangian can be rewritten in terms of $S_{\mu\nu}$ and $F_{\mu\nu}$ as

\begin{align}
 S_{GP} = \int d^4 x \sqrt{-g}({\cal L}_2 + {\cal L}_3 +{\cal L}_4),
\end{align}
with
\begin{subequations}
\ba
{\Lag_2} &=& {G_2 + \alpha_6 F_{\mu\nu}F^{\mu\nu} + \alpha_7 A^{\mu}A^{\nu} F_{\mu\rho}F^{\rho}{}_{\nu}
+\alpha_9 F_{\mu\nu}\tilde{F}^{\mu\nu}},\\
{\Lag_3} &=& G_3 \nabla_{\mu}A^{\mu},
\ea
 and
\ba
\Lag_4
&=&G_4 R -2 G_{4,Y} \( (\nabla_\mu A^\mu)^2 -\nabla_\mu A_\nu \nabla^\nu A^\mu \) \notag\\
&=& G_4 R -{1\over 2} G_{4,Y} \(F_{\mu\nu}F^{\mu\nu} + S^\mu_{~\mu}S^\nu_{~\nu}-S_{\mu\nu}S^{\mu\nu} \)\,,
\ea
\end{subequations}
 where $K, \alpha_6, \alpha_7, \alpha_9, G_3$ and $G_4$ are functions of $Y$.
The second term in the last line can be removed by the redefinition of 
{$\alpha_6$}.
This Lagrangian is found when 
\ba
f = G_4, \quad \alpha_1=-\alpha_2=2G_{4,Y}, \quad 
\alpha_3=\alpha_4=\alpha_5=\alpha_8=0.
\label{caseA}
\ea
Here, $\alpha_6$, $\alpha_7$, and $\alpha_9$ will be still chosen as free parameters.
In this case, the kinetic matrix (\ref{kineticM}) can be written as
\ba
\M=
\left(
\begin{array}{cccc}
	~0 & 0 & 0 & 0 \\[1.5ex]
	~0 &~~\dfrac{(\beta -4 f_Y)}{4}~~ & ~0~ & ~~-2 \sqrt{2} f_Y |\Ah|~ \\[1.5ex]
	~0 & 0 & 0 & ~~ -\sqrt{2} (f-2 Y f_Y)~ \\[1.5ex]
	~0 & ~\ast~ & ~\ast~ & ~~-2 {A}_*^2 f_Y-f ~\\
\end{array}
\right),
\ea
where asterisks represent symmetric components. 
One observes that all the first row and column are zero, leading to $\det \M=0$. 
This ensures that the GP theory is the degenerate theory, and one can freely choose $\beta$ {and $\alpha_9$} as stated in the above.

One also needs to check the other determinants, and they are in fact non-zero in general, 
\begin{subequations}
\ba
\det \M_1&=&\frac{1}{16Y^2}\[2Y \Bigl( f(2f_Y + \alpha_6)-2f_Y Y \alpha_6 \Bigr)
+ \Bigl( 8f_Y^2 Y + f(2\alpha_6+\beta)-2f_Y Y (2\alpha_6 +\beta) \Bigr)\As^2 \]^2, \notag\\
&&\\
y_1&=& -{1 \over 2} (4f_Y-\beta)(f-2f_Y Y)^2,\\
y_2&=& -2 (f - 2f_Y Y)^2 + 4 f_Y (f - 2 f_Y Y) - \frac{1}{4}f (12 f_Y + \beta) - \frac{1}{2}f_Y (12 f_Y + \beta) A_{*}^2, \\
y_3&=& f+f_Y - {\beta \over 4} + 2f_Y \As^2.
\ea
\end{subequations}
Note that if $\beta=4f_Y$ or $f= c \sqrt{|Y|}$ with a constant $c$, $y_1$ vanishes.
In the former case, $y_2$ and $y_3$ does not vanish for 
any arbitrary functions with $f \neq 0$. 
Furthermore, the vector sector degenerate once we impose the additional condition 
$\alpha_6=-2f f_Y/(f-2f_Y Y)$. 
On the other hand, 
in the latter case $f=c\sqrt{|Y|}$, $y_2$ can also vanish
if we make an additional choice of the arbitrary function, $\beta=-6c\sqrt{|Y|}/Y$.
In this case, $y_3 \neq 0$.
Theories with the above parameter choices might imply the existence of an additional primary constraint, 	which can eliminate one of the three vector degrees of freedom.

\subsection{Beyond generalized Proca theory}
The beyond GP theory \cite{2016arXiv160505565H} up to the quartic Lagrangian can be written as
\begin{align}
 S_{BGP} = S_{GP} + \int d^4 x \sqrt{-g}\, {\cal L}_4^{\rm (B)},
\end{align}
with
\ba
\Lag_4^{\rm ({B})}
&=& G_4^{\rm ({B})} \eps^{\mu\nu\rho\sigma}\eps^{\alpha\beta\gamma}_{~~~~\sigma} A_\mu A_\alpha \nabla_\nu A_\beta  \nabla_\rho A_\gamma\notag\\
&=&{1\over 4} G_4^{\rm ({B})} 
\(
2A^\alpha A^\beta F_{\alpha}^{~\gamma} F_{\beta\gamma}
-A_\alpha A^\alpha F_{\beta\gamma} F^{\beta\gamma}
-2A^\alpha A^\beta S_\alpha^{~\gamma}S_{\beta\gamma} \notag\\
&& \qquad \qquad \qquad
+A_\alpha A^\alpha S_{\beta\gamma}S^{\beta\gamma}
+2A^\alpha A^\beta S_{\alpha\beta}S^\gamma_{~\gamma}
-A_\alpha A^\alpha S^\beta_{~\beta}S^\gamma_{~\gamma}
\) \,.
\label{L4beyond1}
\ea
Note that the first and second terms can be absorbed into $\Lag_2$.
One can construct another interaction from the beyond Horndeski 
theory via replacements such as $\nabla_\mu \phi \to A_\mu$\footnote{
 	In fact, three types of the quartic beyond GP Lagrangian can be constructed by using the
 	replacement $\nabla_\mu \phi \to A_\mu$, which are (\ref{L4beyond1}),  (\ref{L4beyond2}), and ${\tilde G}_4^{\rm (beyond)} \eps^{\mu\nu\rho\sigma}\eps^{\alpha\beta\gamma}{}_\sigma A_\mu A_\alpha S_{\nu\beta}  F_{\rho\gamma}$. However, the last interaction is trivially zero, and this is the reason why the term with $\alpha_8$ is missing in \cite{2016PhLB..757..405B}.},
\ba
\Lag_{4,2}^{\rm ({B})}
&=& {\tilde G}_4^{\rm ({B})} \eps^{\mu\nu\rho\sigma}\eps^{\alpha\beta\gamma}_{~~~~\sigma} A_\mu A_\alpha \nabla_\nu A_\rho  \nabla_\beta A_\gamma \notag\\
&=&{1\over 2} {\tilde G}_4^{\rm ({B})} 
\(
2A^\alpha A^\beta F_{\alpha}^{~\gamma} F_{\beta\gamma}
-A_\alpha A^\alpha F_{\beta\gamma} F^{\beta\gamma}
\) \,.
\label{L4beyond2}
\ea
But one will immediately notice that this is already included in $\Lag_2$.
Therefore, the beyond GP theory
is given by
\ba
&&f = G_4, \quad \alpha_1=-\alpha_2=2G_{4,Y}+G_4^{\rm ({B})} Y,\quad\alpha_3=-\alpha_4=2 G_4^{\rm ({B})}, \quad \alpha_5=\alpha_8=0.
\label{caseB}
\ea
Again, $\alpha_6$, $\alpha_7$, and $\alpha_9$ can be still chosen as free parameters.
In this case, the kinetic matrix is explicitly written as 
\ba
\M=
\left(
\begin{array}{cccc}
	~0 & 0 & ~0~ & \dfrac{\alpha _4 {A}_* \Ah^2}{ \sqrt{2}} \\
	~0 & ~~\dfrac{(\beta -4 f_Y)}{2}~~ & ~0~ & - \dfrac{(\alpha _4 {A}_*^2+4f_Y)|\Ah|}{ \sqrt{2}} \\
	~0 & 0 & 0 &~~ \dfrac {4 Y f_Y-2 f+\alpha _4 ({A}_*^4+2{A}_*^2 Y)}{\sqrt{2}} ~\\
	~\ast  &  \ast  &  ~\ast~ &\dfrac{1}{2} {A}_*^2 (\alpha _4 Y-4 f_Y)-f \\
\end{array}
\right),~~
\ea
where asterisks again represent symmetric components. 
Since the first row is linearly dependent on the third row, the determinant of the kinetic matrix vanishes, and hence the beyond GP theory is also the degenerate theory.
Then, $\det \M_1$, $y_1$, $y_2$, and $y_3$ are given by
\begin{subequations}
\begin{align}
& \det \M_1=\frac{1}{64 Y^2}
\Bigl[ 2 Y \Bigl( 2 f \alpha_6 + (f - \alpha_6 Y) (4 f_Y-\alpha_4 Y) \Bigr) \notag\\
&~~~~~~~~~~~~~~~~~~~~~~
 + \Bigl( - 2 (f + 2 Y f_Y) \alpha_4 Y +16 Y f_Y^2
 + (2f - 4 Y f_Y + \alpha _4 Y^2) (2\alpha _6+\beta ) \Bigr) {A}_*^2 \Bigr]^2, \qquad \displaybreak[1]\\
&y_1=\frac{1}{8} (\beta-4f_Y)
\[
4(f-2f_Y Y)^2 -Y\alpha_4 (8f-16 f_Y Y-\alpha_4 Y)\As^2 \notag\\
&~~~~~~~~~~~~~~~~~~~~~~~
- 2\alpha_4 (2f-4f_Y Y-\alpha_4 Y -2\alpha_4 Y^2) \As^4+(1+4Y)\alpha_4^2 \As^6 + \alpha_4^2 \As^8
\],\displaybreak[1]\\
&y_2=-2f^2 -{\beta \over 4} f + f f_Y(1+8Y) -8f_Y^2 Y(1+Y)\notag\\
&~~~~~~~~
+{1 \over 8} \[ 32Y\alpha_4 f-Y\alpha_4(4Y\alpha_4-\beta) -48f_Y^2-4f_Y \(Y(9+16Y)\alpha_4+\beta \) \] \As^2\notag\\
&~~~~~~~~
-{1\over 2} \alpha_4 \(Y(3+4Y)\alpha_4-4f+8f_Y(1+Y)\)\As^4
-(1+2Y)\alpha_4^2 \As^6 -{\alpha_4^2 \over 2}\As^8,\displaybreak[1]\\
&y_3=f+f_Y-{\beta \over 4} +\(2f_Y-{Y\alpha_4 \over 2}\)\As^2.
\end{align}
\end{subequations}

One might notice that $y_1$ can be zero only if $\beta=4f_Y$ with non-zero $\alpha_4$, 
and the vector sector again degenerate when  
$\alpha_6=f(\alpha_4 Y -4f_Y)/(2f+(\alpha_4 Y -4 f_Y)Y)$.
As in the previous case, this choice might eliminate an additional degrees of freedom in vector modes, however this should be carefully checked in Hamiltonian analysis, which will not be discussed in the present paper.
As investigated in appendix \ref{appendix-trf.}, 
the beyond GP theory can be obtained from the GP theory by performing a disformal transformation, and the disformal factor $\Gamma(Y)$ therefore corresponds to the additional parameter $G_4^{(B)}(Y)$ in the beyond GP theory. 
On the other hand, the theory obtained from the beyond GP theory by performing a disformal transformation still belongs to the beyond GP theory up to the redefinition of arbitrary functions.
However, once we perform a conformal transformation and/or vector field redefinition from the GP theory, the resultant theory no longer belongs to the beyond GP theory. 

\subsection{Case A : General theories with $\alpha_1+\alpha_2=0$} \label{subsec:CaseA}

In this subsection, we consider the case $\alpha_1+\alpha_2=0$, which is required to satisfy the first degeneracy condition (\ref{degeneracy}), and here assume 
	$f\neq 0$. The special case where the Einstein-Hilbert term is absent is investigated in appendix \ref{appendix-CaseC}.
First, let us consider (\ref{eq-DY3}),
which is also required to vanish. 
Substituting $\alpha_1=-\alpha_2$ in (\ref{eq-DY3}), 
we obtain
	\ba
	0 &=&-{1 \over 16} (f + \alpha_2 Y) \( (\alpha_4+\alpha_8)Y - 2\alpha_2 -\beta\)\notag\\
	&&\times
	\[8 (\alpha _3+\alpha _4 +\alpha _5 Y) (f+ \alpha _2 Y)
	 - 3 (2 \alpha _2+4 f_Y+\alpha _3 Y){}^2 \] .
	\ea
This has three branches, which are given by
\ba
&&\alpha_2=-{f\over Y}
\quad  {\rm or}\quad
\alpha_2={(\alpha_4+\alpha_8)Y-\beta\over 2} 
\quad \nonumber\\
&& {\rm or}\quad
\alpha_4=
\frac{3 (2 \alpha _2+4 f_Y +  \alpha _3 Y)^2}{8 (f+\alpha
	_2 Y)} - (\alpha_3 + \alpha _5 Y).
\label{C1}
\ea
Here, we assumed $f+\alpha_2 Y \neq 0$ in the last branch of (\ref{C1}).
Substituting the first branch of (\ref{C1}) into $D_1(Y)$, we get 
\ba
0 = D_1(Y) = \frac{f}{16 Y^3} \Bigl( Y (4 f_Y+\alpha _3 Y)-2 f \Bigr)^2
 \Bigl( 6 f+Y \beta -(\alpha _4+\alpha _8) Y^2 \Bigr) .
\ea
Then, we have two solutions in the first branch of (\ref{C1}), \\
{\bf  \underline{Case~A1} :}
\ba
&&
\alpha_1=-\alpha_2={f\over Y}, \quad \alpha_3=\frac{2(f-2f_Y Y)}{Y^2}, 
\label{caseA1}
\ea
{\bf  \underline{Case~A2} :}
\ba
&&
\alpha_1=-\alpha_2={f\over Y}, \quad \alpha_4=\frac{6f+Y\beta}{Y^2}-\alpha_8.
\label{caseA2}
\ea
In both cases, each matrix element of $\M$ is not zero.
The case A$1$ corresponds to the "class Ib" in the scalar-tensor theories, which was found in \cite{2016arXiv160208398B}. 
This is because the degeneracy condition (\ref{caseA1}) is completely independent on $\beta$ and $\alpha_8$, that is, 
	only the symmetric part of $\nabla_\mu A_\nu$ in the Lagrangian (\ref{action}) plays a role to satisfy the degeneracy condition (\ref{degeneracy}). 
	As confirmed in \cite{2016arXiv160208398B}, the vector sector further degenerate in the "class Ib". 
	On the other hand, in our case, the vector sector does not still degenerate since we now have two additional vector components, ${\dot A}_2$ and ${\dot A}_3$, in the matrix $\M_1$, which are apparently absent in the scalar-tensor theories. 

Substituting the second branch of (\ref{C1}) into $D_1(Y)$, we get 
\ba
0 = D_1(Y)
= \frac{1}{32} Y \Bigl[ \((\alpha _4+\alpha _8) Y-\beta
\) (8 f_Y-2 \beta+\alpha _8 Y) - 2 (2\alpha_3+2 \alpha _4 + \alpha _8) f \Bigr]^2.
\ea
One can easily solve this for $\alpha_3$, and we have only one solution,\\
{\bf  \underline{Case~A3} :}
\ba
&&
\alpha_1=-\alpha_2=-{(\alpha_4+\alpha_8)Y-\beta\over 2} , \notag\\
&& \alpha_3=\frac{\bigl((\alpha _4+\alpha _8) Y-\beta \bigr) (-2
	\beta +8 f_Y+\alpha _8 Y)-2 (2 \alpha _4+\alpha
	_8) f}{4 f}.
\label{caseA3}
\ea

$D_1(Y)$ in the third branch of (\ref{C1}) can be written as the power series in $\alpha_8$ as in Eq.~(\ref{d1d0}), 
\ba
 0 &=& D_1(Y)
= W_1(f, \alpha_2) \, \alpha_8^2 + W_2(f, \alpha_2, \alpha_3
)\, \alpha_8 + W_3 (f, \alpha_2, \alpha_3, \alpha_5, \beta) \,, 
 \label{C1dD1}
\ea
where $W_1$, $W_2$, and $W_3$ are given by
\begin{subequations}
\ba
W_1(f, \alpha_2)&=&
\frac{1}{8} Y (f+\alpha _2 Y){}^2,\\
W_2(f, \alpha_2, \alpha_3
)&=&
\frac{1}{16} Y (2 \alpha _2+4 f_Y+\alpha _3 Y)
 \biggl[4 f_Y (4 \alpha _2 Y + 3 f) -f(2 \alpha _2+\alpha _3 Y) \biggr], \,\,\,\qquad \\
W_3(f, \alpha_2, \alpha_3, \alpha_5, \beta)&=&
{\cal P}_1(f, \alpha_2, \alpha_3, \beta) \alpha_5+ {\cal P}_2(f, \alpha_2, \alpha_3, \beta) (2 \alpha _2+4 f_Y+\alpha _3 Y),
\ea
\end{subequations}
with
\begin{subequations}
\begin{align}
{\cal P}_1(f, \alpha_2, \alpha_3, \beta) &=
 -\frac{1}{16} Y  \[ 8 \alpha _2 Y^2 (\alpha _2 \beta +2\alpha _3 Y f_Y+8 f_Y^2)+8f^2 (2 \alpha _2+\beta +\alpha _3 Y)\nonumber\\
&~~~~~~~~~~~~~~~~~~~~~~~~
+Y f \(16 \alpha _2 \beta +12 \alpha _2^2+8 f_Y(\alpha _3 Y-2 \alpha _2)
\nonumber\\
&~~~~~~~~~~~~~~~~~~~~~~~~
+48f_Y^2-\alpha _3^2 Y^2+4 \alpha _2 \alpha _3Y \) \],\displaybreak[1]\\
{\cal P}_2(f, \alpha_2, \alpha_3, \beta) &={1\over 128} Y (f+\alpha _2 Y)^{-1}\notag\\
&~~~
\times\[
32 \alpha _3 f^2 (2 \alpha _2+\beta +\alpha _3Y)
\nonumber\\
&~~~~~~~~~
+f \(-24 \alpha _2^3-4 f_Y (8\alpha _2 \beta +36 \alpha _2^2-3 \alpha _3^2Y^2+44 \alpha _3 \alpha _2 Y)\nonumber\\
&~~~~~~~~~
+48 f_Y^2(2 \alpha _2+5 \alpha _3 Y)+576f_Y^3-3 \alpha _3^3 Y^3+14 \alpha _2 \alpha _3 Y(4 \beta +\alpha _3 Y)\nonumber\\
&~~~~~~~~~
+4 \alpha _2^2(7 \alpha _3 Y-4 \beta )\)+8
\alpha _2 Y \(\alpha _2 \beta  (-2\alpha _2-4 f_Y+3 \alpha _3 Y)\nonumber\\
&~~~~~~~~~
+2 f_Y\bigl(24 f_Y (\alpha _2+\alpha _3
Y)+48 f_Y^2+\alpha _3 Y (3 \alpha _3Y-2 \alpha _2)\bigr) \)
\].
\end{align}
\end{subequations}
Therefore, one can solve this for $\alpha_8$
\footnote{Although the quadratic equation in $\alpha_8$ (\ref{C1dD1}) has two solutions,
 we unified two branches into one case
 in our classification because $+(-)$ branch is connected with $-(+)$ branch via a metric transformation (\ref{gtilde}) and a field redefinition (\ref{Atilde}).}
and we have \\
{\bf  \underline{Case~A4} :}
\ba
\alpha_1=-\alpha_2 \,, \quad
 \alpha_4= \frac{3 (2 \alpha _2+4 f_Y +  \alpha _3 Y)^2}{8 (f+\alpha
	_2 Y)} - \alpha_3 -\alpha _5 Y \,, \quad
 \alpha_8=\frac{-W_2 \pm \sqrt{W_2^2 -4W_1W_3}}{2W_1}.\label{C1d}
\label{caseA4}
v\ea
Since $\alpha_8$ has to be a real function, we require $W_2^2 -4W_1W_3 \geq 0$, and $f$ and $\alpha_{2, 3, 5}$ have to be chosen so that this condition is satisfied.
In the above analysis, we have assumed $f+\alpha_2Y \neq 0$ and the case $f+\alpha_2Y = 0$ is included in the first branch, namely the case A$1$ or A$2$.

\subsubsection{Relation with the (beyond) generalized Proca theories}
\label{trfA4}

Here, we discuss the relation of the case A with the GP and beyond GP theories.
As we will see below, a certain subclass of these theories are included in the cases A$1$, A$2$ and A$3$.
However, as is clear from the number of the arbitrary functions, 
new theories are also included in these cases.
On the other hand, all the remaining theories of the GP and beyond GP theories belong to the case A$4$.
More interestingly, all the rest of the case A$4$ is nothing but the one that can be obtained through 
the transformations (\ref{gtilde}) and (\ref{Atilde}) of the GP and beyond GP theories.
This is compatible with the fact that the number of arbitrary functions in the case A$4$ is equal to that in the GP theory plus three transformation parameters,
that is, the conformal, the disformal and the rescaling factors\footnote{In the case of the beyond GP theory, the transformation parameters are only the conformal and the rescaling factors since its action is invariant under the disformal transformation up to the redefinition of arbitrary functions.
However, the number of arbitrary functions in the case A$4$ is equal to the number of arbitrary functions ($G_4$, $G_4^{(B)}$, $\alpha_6$ and $\alpha_7$) plus two transformation parameters as in the case of the GP theory.}.
However, new theories also exist in the case A4 as we see below.

First, we make clear the relation of the cases A$1$, A$2$ and A$3$ with the GP and beyond GP theories.
Comparing the parameters in the case A$1$ (\ref{caseA1}) with those in the GP theory (\ref{caseA}),
the case A$1$ includes the special case of the GP theory when 
\ba 
  f = G_4 = c \sqrt{|Y|} \,, \quad
  \alpha_4 = \alpha_5 = \alpha_8 = 0 \,,
  \label{corA1}
\ea
 where $c$ represents an arbitrary constant.
On the other hand, the special case of the beyond GP theory can be included in the case A$1$ when
\ba 
  f = G_4 \,, \quad
  \alpha_4 = - \alpha_3 \,, \quad
  \alpha_5 = \alpha_8 = 0 \,, \quad
  G_4^{(B)} = \frac{G_4 - 2 Y G_{4, Y}}{Y^2} \,.
\ea
As for the case A$2$, one notice that a subclass of the GP theory is included when
\ba 
  f = 
G_4 = c \sqrt{|Y|} \,, \quad
  \alpha_3 = \alpha_5 = \alpha_8 = 0 \,, \quad
  \beta = - \frac{6 G_4}{Y} 
= - 6c \frac{\sqrt{|Y|}}{Y}\,.
  \label{corA2}
\ea
Now the arbitrary function $\beta$ must be also tuned when $f \neq 0$.
A subclass of the beyond GP theory belongs to this case if 
\ba 
 f = G_4 \,, \quad
 \alpha_3 = - \alpha_4 \,, \quad
 \alpha_5 = \alpha_8 = 0 \,, \quad
G_4^{(B)} = \frac{G_4 - 2 Y G_{4, Y}}{Y^2} \,, \quad
 \beta = 4 \frac{Y G_{4, Y} - 2 G_4}{Y} \,.
\ea
In the case A$3$,
the GP theory is included when
\ba 
 f = G_4 \,, \quad
  \alpha_4 = \alpha_5 = \alpha_8 = 0 \,, \quad
\beta = 4 G_{4, Y} \,,
  \label{corA3}
\ea
 and 
the beyond GP theory is included when
\ba 
  f = G_4 \,, \quad
  \alpha_4 = -2 G_4^{(B)} \,, \quad
  \alpha_5 = \alpha_8 = 0 \,, \quad
  \beta = 4 G_{4, Y} \,.
 \label{corA3B}
\ea
The parameter choices $f=c \sqrt{|Y|}$
	in (\ref{corA1}) and (\ref{corA2}) and $\beta = 4 G_{4, Y}$ in (\ref{corA3}) and  (\ref{corA3B}) correspond to the cases that the scalar sector is further degenerate, $y_1=0$.

As we have seen, a subclass of the GP and beyond GP theories with particular arbitrary functions
 are included in the cases A$1$, A$2$ and A$3$, but there must exist new theories in those cases
 according to the number of the arbitrary functions.
For example in the case A$1$, we have arbitrary functions, $\{f, \alpha_4, \alpha_5, \alpha_6, \alpha_7, \alpha_8\}$\footnote{
	One can choose different free functions, for example, $\alpha_1$ instead of $f$ in the case A1 (\ref{caseA1}). In this case, the set of free functions is given by  $\{\alpha_1, \alpha_4, \alpha_5, \alpha_6, \alpha_7, \alpha_8\}$, however the number of free functions is indeed unchanged.
	}  while the GP theory with $G_4 \propto \sqrt{|Y|}$ and an arbitrary $\beta$ belongs to this case.
Given the fact that we only have three transformations factors \eqref{gtilde} and \eqref{Atilde}, apparently
 we cannot explore the whole parameter space through the transformation of the above specific GP theory.
Then, we conclude that new degenerate vector-tensor theories
must exist in the case A$1$ as well as the cases A$2$ and A$3$,
which cannot be obtained from any known theories through transformations.
Furthermore, we confirmed that the theories after the transformations still satisfy the same degeneracy conditions of the original frame, and different branches based on our classification are never mixed
 by any kinds of the transformations that we have introduced in the present paper.

\begin{figure}[t]
	\begin{center}
		\includegraphics[width=120mm]{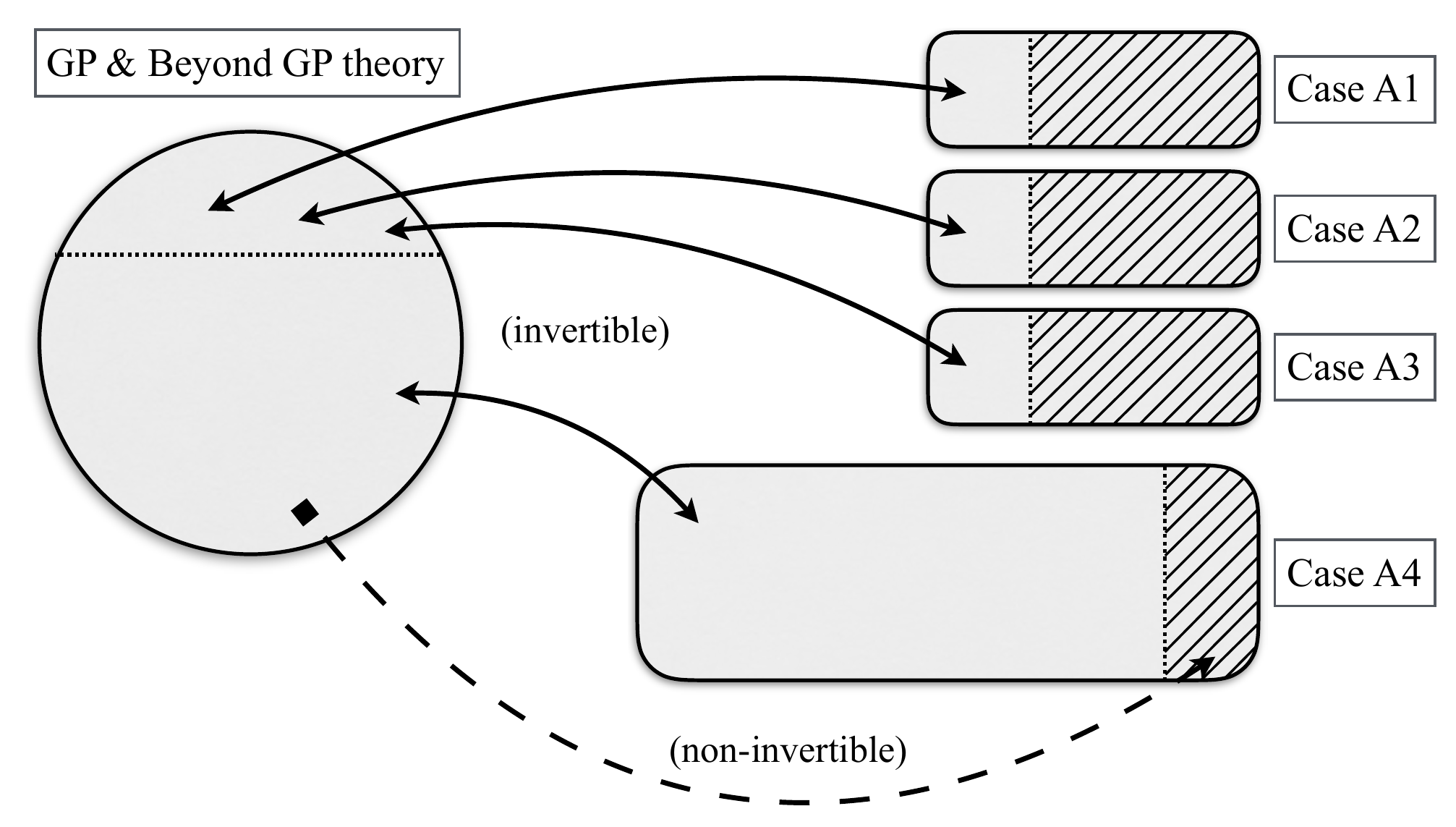}
	\end{center}
	\caption{The relation between the (beyond) GP theories and the case A through the transformations (\ref{gtilde}) and (\ref{Atilde}). The shaded regions correspond to the new theories of massive vector field in curved space-time, which satisfies the degeneracy condition (\ref{degeneracy}). The arrow with the straight line represents invertible transformation, and the dashed arrow represents non-invertible transformation.
		}
		\label{figure2}
	\end{figure}

Let us move to the case A$4$.
The degeneracy conditions (\ref{C1d}) is satisfied with the parameters of the GP (\ref{caseA}) and the beyond GP (\ref{caseB}) theory except for special cases (\ref{corA1})-(\ref{corA3B}), which are related with the cases A1, A2 and A3.	
Therefore, one concludes that 
both the GP and the beyond GP theories are included in the case A4 in general.
The degeneracy conditions (\ref{C1d}) are strictly preserved by the metric transformation (\ref{gtilde}) and the field redefinition (\ref{Atilde}). Thus, any theory transformed from the GP theory, including the beyond GP theory, belongs to the case A4. 
Moreover, since there are six arbitrary functions $\{f, \alpha_2, \alpha_3, \alpha_5, \alpha_6, \alpha_7\}$
 in the case A4, one would expect that the GP theory can be mapped from the case A4. To see this, we want to re-express six parameters, namely three arbitrary functions in the GP theory, $\{G_4, {\bar \alpha_6}, {\bar \alpha_7} \}$, and three transformation parameters, $\{ \Omega, \Gamma,  \Upsilon \}$,
  in terms of six functions in the case A4, $\{f, \alpha_2, \alpha_3, \alpha_5, \alpha_6, \alpha_7\}$. 
Using (\ref{TGP}), we get the first order differential equations for $\Omega$ and $\Gamma$ as
\begin{subequations}
\ba
\Omega_Y&=&  \frac{4f_Y +2\alpha_2+Y\alpha_3}{4(f+\alpha_2 Y)}\Omega,\label{omegaY}\\
\Gamma_Y&=& 2 \frac{\Gamma  f+\alpha _2 (\Gamma Y+\Omega )}{f+\alpha _2 Y} \frac{\Upsilon _Y}{\Upsilon} + \frac{2 \alpha _2 (\alpha _2+2 f_Y)-\alpha _3 (2 f+\alpha _2 Y)}{4 (f+\alpha _2 Y){}^2}
  \Omega \,.
\label{gammaY}
\ea
\end{subequations}
Here, we have used the condition of the case A$4$, $f+Y\alpha_2 \neq 0$.
On the other hand, $\Upsilon$ is determined by the following first order differential equation,
\ba
Z_1 \left({\Upsilon_Y\over \Upsilon}\right)^2+2YZ_2 \left({\Upsilon_Y\over \Upsilon}\right)+Z_2 =0,
\ea
where we defined 
\ba
Z_1&=&(f+Y\alpha_2)\[
8 (f + Y \alpha_2) (Y^2 \alpha_5 - \beta - 2 \alpha _2 - Y \alpha _3) \notag\\
&& \qquad \qquad \qquad \qquad
+ Y (2\alpha_2 + 4 f_Y + \alpha_3 Y) \(10 \alpha _2 - 3 (4 f_Y+\alpha _3Y) \) \],\\
Z_2&=& 8 \alpha_5 (f + Y \alpha_2)^2
+ (2 \alpha _2+4 f_Y+\alpha _3 Y) \(2 \alpha _2 (\alpha _2+2 f_Y)-\alpha _3 (4 f+3\alpha _2 Y)\) .
\ea
One can easily solve the equation for $\Upsilon$, and we get 
\ba
{\Upsilon_Y\over \Upsilon}
= \left\{ 
\begin{array}{cr}
	\displaystyle{
	\frac{-YZ_2 \pm  \sqrt{Z_2 (Y^2 Z_2 - Z_1)}}{Z_1}
	} 
	&{(Z_1 \neq 0)}, \\
	\displaystyle{}\\
	\displaystyle{
	-{1\over 2Y}
	} 
	&{(Z_1 = 0{~\rm and~} Z_2\neq 0)}.\\
\end{array} \right.\label{upsilonY}
\ea
Surprisingly, the expression inside the square root in the solution of $\Upsilon_Y/\Upsilon$, (\ref{upsilonY}), is proportional to the one in the solution of $\alpha_8$, (\ref{caseA4}), namely 
\ba
Z_2 (Y^2 Z_2 - Z_1) =\frac{256}{Y^2} (W_2^2-4W_1W_3),
\ea
which is required to be positive in the case A4.
Therefore, we do not have to impose any additional constraint for the parameters.
Hence one can always relate 
$\{G_4, {\bar \alpha_6}, {\bar \alpha_7}, \Omega, \Gamma,  \Upsilon\}$ and $\{f, \alpha_2, \alpha_3, \alpha_5, \alpha_6, \alpha_7\}$ by using the integrated expression of (\ref{omegaY}), (\ref{gammaY}), and (\ref{upsilonY}). In other words, 
the case A4 can be always mapped to the GP theory.
 After solving those differential equations one can also express the remaining functions as 
 \begin{subequations}
 \begin{align}
G_4&=\frac{f}{\sqrt{\Omega(\Omega+Y\Gamma)}},\displaybreak[1]\\
{\bar \alpha_6} &=\frac{f \Gamma + \Omega \alpha_6}{\Upsilon^2\sqrt{\Omega(\Omega+Y\Gamma)}}, \displaybreak[1]\\
{\bar \alpha_7} &=
 \frac{\sqrt{\Gamma  Y+\Omega }}{\Upsilon^2 Y \Omega^{5/2} \left(\Upsilon +Y \Upsilon _Y\right){}^2}
\[
4 Y \frac{\Upsilon_Y}{\Upsilon} \Omega \( \Gamma ^2 Y (f+\alpha _2 Y)
+2  \alpha _2 \Gamma Y \Omega
+ (\alpha _2-\alpha _6) \Omega ^2 \) \notag\\
&~~~~~~~~~~~~~~~~~~~~~~~~~~~~~~~
+ 2 Y \Omega _Y^2 f (\Gamma  Y + \Omega)
+ 2 Y^2 \frac{\Upsilon_Y^2}{\Upsilon^2} \Omega 
 \( \Gamma^2 Y f + \alpha _2 (\Gamma Y + \Omega)^2 - \alpha _6 \Omega^2 \)
 \notag\\
&~~~~~~~~~~~~~~~~~~~~~~~~~~~~~~~
+ \Omega \( 2 \alpha_2 (\Gamma Y + \Omega)^2
 + 2 Y \Gamma (\Gamma f + \alpha_6 \Omega) \)\notag\\
&~~~~~~~~~~~~~~~~~~~~~~~~~~~~~~~
+ \Omega^2 (\Gamma Y + \Omega) (4 f_Y+\alpha _7 Y) 
- 4 \Omega \Omega _Y (\Gamma  Y + \Omega) (2 Yf_Y+f) \].
 \end{align}
\end{subequations}
In figure \ref{figure2}, we show the schematic figure of the relation between the (beyond) GP theories and the case A via the transformations (\ref{gtilde}) and (\ref{Atilde}). 

Before the end of this section, 
	let us make a comment on the case with non-invertible transformation.
As studied in \cite{Chamseddine:2013kea,Deruelle:2014zza},
if the transformation is non-invertible,
where the inverse transformation does not exist, 
 the resultant theory has nothing to do with the original theory. 
The special cases in the case A$4$, which can be formally mapped from the GP theories through such non-invertible transformations, have completely different properties from those in the corresponding GP theories. 
Such a case can be for example found when the parameters in the case A$4$ satisfy
\ba
 && f = Y \,, \qquad
 \alpha_3 = \frac{2 \alpha_2}{Y} \,, \qquad
 \alpha_5 = \frac{2 \alpha_2}{Y^2} \,,  \qquad
\label{singularA4}
\ea
Further careful study is necessary for such special cases.
We will defer this interesting subject in future work.

\subsection{Case B : General theories with $\alpha_1 + \alpha_2 \neq 0$
}
\label{subsec:CaseB}
We now consider the second branch of (\ref{caseC}) where $Q(f, \alpha_1, \alpha_2, \alpha_4, \alpha_8, \beta)=0$
 while $\alpha_1 + \alpha_2 \neq 0$. For simplicity, we here set $f=1$ and $\beta=0$ throughout this subsection. As investigated in appendix \ref{appendix-trf.}, these parameter choices can be realized after performing the conformal and the field redefinition without loss of generality when $f \neq 0$, for completeness\footnote{Here, we implicitly assumed $f>0$. One can in general consider $f<0$, however it might lead to a wrong sign of the kinetic or gradient term of the tensor mode. Such an example can be found in the cosmological solution of the GP theory in \cite{2016arXiv160505066D}, and the tensor perturbation suffers from the ghost/gradient instabilities when $f<0$.~For $f>0$, $f$ can be always set to be unity by a transformation with $\Omega>0$.}.
The case $f = 0$ is individually investigated in appendix \ref{appendix-CaseC}.

As investigated in appendix \ref{appendix-trf.},  the condition $\alpha_1 + \alpha_2 \neq 0$ is preserved under any of conformal transformation, disformal transformation (\ref{gtilde}), and vector field redefinition (\ref{Atilde}). 
In other words, any theories in the case B are not connected to the GP and the beyond GP theories where $\alpha_1+\alpha_2=0$ under the metric transformation and the vector field redefinition, and the case B can be therefore categorized as the new class of theories.

After plugging $f=1$ and $f_Y=\beta=0$ into the second branch of (\ref{caseC}), we have
 \footnote{
It should be noted that, if $\alpha_8 = 4/ Y^2$, the equation (\ref{D0C2}) completely coincides with the corresponding equation in the scalar-tensor theory. 
 }
\ba
Q = 8\(2\alpha_1+Y(\alpha_4-\alpha_8)\) +Y^3\alpha_8^2=0.\label{D0C2}
\ea
 In this case, we can solve this for $\alpha_4$, and we have
\ba
\alpha_4=
-{2\alpha_1 \over Y} + \alpha_8 - {Y^2\alpha_8^2 \over 8}.
\label{eq-a4B}
\ea
Again, we consider (\ref{eq-DY3}) by utilizing $f = 1, \beta = 0$ and (\ref{eq-a4B}), which now reduces to
\ba
0 
&=& -{1\over 256}\alpha_8(1-\alpha_1 Y) (16-\alpha_8 Y^2)\notag\\
&&\times \[
\(2+(\alpha_1+3\alpha_2)Y\) \(8\alpha_8 Y - \alpha_8^2 Y^3 -8(\alpha_1-\alpha_2) +8Y(\alpha_3+Y\alpha_5)\)
-6Y(2\alpha_2+\alpha_3 Y)^2
 \] \,.\notag\\ 
\ea
Then, we have five branches, which are given by
\ba
&&\alpha_8=0
\quad  {\rm or}\quad
\alpha_1={1\over Y}
\quad  {\rm or}\quad
\alpha_8={16 \over Y^2},
\quad  {\rm or}\notag\\
&& 
\alpha_5=
\frac{\alpha_8^2Y}{8}-\frac{\alpha_8}{Y}
+ \frac{8 (\alpha_1 - \alpha_2) + 4 (\alpha_1^2 + 2 \alpha_1 \alpha_2 - 2 \alpha_3) Y
	- 4 \alpha_1 \alpha_3 Y^2 + 3 \alpha_3^2 Y^3}{4 Y^2 \( 2 + (\alpha _1+3 \alpha_2) Y \)},
\notag\\
&& {\rm or}\quad
\alpha_2=-{2+Y\alpha_1 \over 3Y}.
\label{C2}
\ea
Here, we assume that  degeneracy conditions in the former branch of (\ref{C2}) is non-zero, for example, 
	$\alpha_8 \neq 0$ for the second branch $\alpha_1=1/Y$.
The first branch of (\ref{C2}) gives
\ba
 0 = D_1(Y)=\frac{(2\alpha_1-Y\alpha_3)^2}{2Y}-2Y(\alpha_1+\alpha_2)\alpha_5 .
\ea
We solve this for $\alpha_5$, and since $\alpha_1+\alpha_2 \neq 0$ we get 
\\
{\bf  \underline{Case~B1} :}
\ba
&&f=1, \quad \beta=0, \quad 
\alpha_4=-{2\alpha_1 \over Y}, \quad
\alpha_5=\frac{(2\alpha_1-Y\alpha_3)^2}{4Y^2 (\alpha_1+\alpha_2)}, \quad \alpha_8=0.
\ea
Now the second branch in (\ref{C2}) gives 
\ba
 0 = D_1(Y)
 = \frac{(\alpha _8 Y^2-8){}^2 \Bigl( 2 (2 - Y^2 \alpha _3)^2
  - Y^2 \alpha_8 (1 + Y \alpha _2) (8 - \alpha _8 Y^2)
  - 8 (1 + Y \alpha _2) \alpha _5 Y^3 \Bigr)}{256 Y^3}.~~
\ea
Thus, we have two solutions, \\
{\bf  \underline{Case~B2} :}
\ba
&&f=1, \quad \beta=0, \quad 
\alpha_1={1\over Y}, \quad \alpha_4=-{2\over Y^2},
\quad \alpha_8={8\over Y^2},
\ea
{\bf  \underline{Case~B3} :}
\ba
&&f=1, \quad \beta=0, \quad 
\alpha_1={1\over Y}, \quad \alpha_4=-{2\over Y^2}+\alpha_8-{Y^2\alpha_8^2 \over 8}\notag\\
&&\alpha_5=\frac{(2-Y^2\alpha_3)^2}{4Y^3(1+Y\alpha_2)}
-{\alpha_8 \over Y}+{Y\alpha_8^2 \over 8},
\label{caseB3}
\ea
where $1+Y\alpha_2 \neq 0$ since $\alpha_1+\alpha_2 \neq 0$.
The third branch in (\ref{C2}) gives 
\ba
 0 = D_1(Y) 
&=& \frac{1}{2 Y^3}\[
  - 4 (\alpha_1+\alpha _2) \alpha _5Y^4
 + \alpha _3^2Y^4 +12 \alpha _1 \alpha _3 Y^3 \notag\\
&& \qquad \qquad
-4 (7\alpha _1^2+16 \alpha _1\alpha _2 +4\alpha _3) Y^2
+32 (\alpha _1+4\alpha _2) Y+64\].
\ea
Solving this for $\alpha_5$ we get\\
{\bf  \underline{Case~B4} :}
\ba
&&f=1, \quad \beta=0, \quad 
\alpha_4=-{16 \over Y^2}-{2\alpha_1 \over Y},\quad \alpha_8={16 \over Y^2},\notag\\
&&\alpha_5=
\frac{\alpha _3^2 Y^4 +12 \alpha _1 \alpha _3 Y^3
  -4 (7\alpha _1^2+16 \alpha _2 \alpha _1 +4\alpha _3) Y^2
 +32 (\alpha _1+4\alpha _2) Y+64}{4Y^4(\alpha_1+\alpha_2)}.
\ea
The fourth branch in (\ref{C2}) gives 
\ba
 0 = D_1(Y)
  = \frac{(1-\alpha _1 Y) \Bigl( 2 \alpha_1 \alpha_8 Y^2 + 8 (2 \alpha _2+\alpha _3Y)
+ Y (4 + 4 Y \alpha_2 - Y^2 \alpha _3) \alpha_8 \Bigr)^2}{64 Y \bigl((\alpha _1+3 \alpha _2) Y+2\bigr)} \,,
\ea
where $(\alpha _1+3 \alpha _2) Y+2 \neq 0$.
The first solution, $\alpha_1=1/Y$, corresponds to the case B3 
since the expression of $\alpha_5$ in the forth branch (\ref{C2}) with $\alpha_1=1/Y$ reduces the one in the case B
	(\ref{caseB3}).
Therefore, we disregard this solution, and we get \\
{\bf  \underline{Case~B5} :}
\ba
&&f=1, \quad \beta=0, \quad 
\alpha_1=\frac{
	-8(2\alpha_2+Y\alpha_3) - Y(4+4Y\alpha_2-Y^2\alpha_3)\alpha_8}{2Y^2\alpha_8}\notag\\
&&\alpha_4={4(1+Y\alpha_2) \over Y^2}-\alpha_3+{8(2\alpha_2+Y\alpha_3) \over Y^3 \alpha_8}+\alpha_8-{Y^2\alpha_8^2 \over 8}\notag\\
&&\alpha_5={-2+Y^2 \alpha_3 \over Y^3}-{4(2\alpha_2+Y\alpha_3) \over Y^4 \alpha_8}-{\alpha_8 \over Y}+{Y\alpha_8^2 \over 8}+{12(2\alpha_2+Y\alpha_3) \over Y^2(Y^2\alpha_8-8)}
\ea
Here, $Y^2 \alpha_8 -8 \neq 0$ since $(\alpha _1+3 \alpha _2) Y+2 \neq 0$.
 The last branch in (\ref{C2}), which corresponds to the case $(\alpha _1+3 \alpha _2) Y+2 = 0$, provides the further condition from (\ref{eq-DY3}),
so let us take a look at this first. It is now given by 
\ba
 0 
=-\frac{\alpha _8 (1-\alpha _1 Y) 	(\alpha _8
	Y^2-16)(-3 \alpha _3 Y^2+2 \alpha _1Y+4){}^2}{384Y}.
\ea
The first solution $\alpha_8=0$ corresponds to the case B1, 
the second solution $\alpha_1=1/Y$ corresponds to the case B3,
and the third solution $\alpha_8=16/Y^2$ corresponds to the case B4.
Therefore, the remaining solution is \\
{\bf  \underline{Case~B6} :}
\ba
&&f=1, \quad \beta=0, \quad 
\alpha_2={-2-Y\alpha_1 \over 3Y}, \quad \alpha_3={2(2+Y\alpha_1 ) \over 3Y^2},\quad
\alpha_4=-{2\alpha_1 \over Y}+\alpha_8-{Y^2\alpha_8^2 \over 8},\notag\\
&&\alpha_5={2(Y\alpha_1-1)\over 3Y^3}-{\alpha_8 \over Y}+{Y \alpha_8^2 \over 8}. \quad
\ea

\section{Summary and discussion}
\label{sec4}
In this paper we propose a new class of extended vector-tensor theories, which carries at most five degrees of freedom, namely three for massive vector mode and two for massless tensor mode. 
Starting from the most general action for vector field 
 which contains up to two derivatives with respect to $g_{\mu \nu}$ and $A_\mu$,
 we have imposed a degeneracy condition on the kinetic part of the action 
 in order to eliminate the would-be Ostrogradski mode.
We then have found a new class of degenerate vector-tensor theories denoted as the cases A, B, and C,
 which are not included in any known theories such as the generalized Proca (GP) and
 the beyond generalized Proca theories. 
We also confirmed that both the GP and beyond GP theories are the degenerate theories even in curved space-time as naively expected in the previous works.

In the course of analysis, we have also extended metric transformations~by incorporating a vector field.
These transformations are characterized by the so-called conformal and disformal factors,
which are functions of the vector field, more rigorously the contraction of vector field, $Y = A_\mu A^\mu$.
In addition, we have introduced a field redefinition of the vector field, to which there is no analog
 in the scalar field language since this is not a mere redefinition of $\phi$ nor $X$.
After checking that the action is invariant under these transformations modulo redefinition of arbitrary functions, we confirmed that classification of such degenerate theories is stable, that is, each case specified in the analysis is never mixed and the same degeneracy condition still holds even after these transformations.
This result clarifies that our vector-tensor theories in the cases A$1$,$~$A$2$ and A$3$, except for the special examples, and all of the case B are new theories, which cannot be obtained by any kinds of the metric transformations and the redefinition of vector field from the known vector-tensor theories such as the GP and the beyond GP theories. 
We found that the remaining branch, the case A4, includes both the GP and the beyond GP theories up to the quartic Lagrangian. Furthermore, the theory which is obtained from the (beyond) GP theory by the transformations (\ref{gtilde}) and (\ref{Atilde}) is also included in the case A4.
Since the number of free functions of this branch is six, these six functions can be regarded as the three free functions of the GP theory and the three free functions of the transformations, as proved in section \ref{trfA4}.
	While the theories which correspond to the non-shaded region in the case A4 in figure \ref{figure2} can be mapped from the GP theory through invertible transformations, and vice versa,
	specific theories which correspond to the shaded region in the case A4 in figure \ref{figure2} are related with the GP theory through non-invertible transformations. 
These specific theories in the case A4 have nothing to do with the GP theory, and hence those theories should be regarded as new theories.
In appendix \ref{appendix-CaseC}, we also classified the new theories with $f=0$ denoted as the case C, 
for completeness.

 One of interesting directions of study is to seek for massless vector theories.
Since the presence of the degeneracy just promises one primary constraint, our vector-tensor theories also include the theories which carry less than five degrees of freedom.
One possibility to remove further degrees of freedom is the presence of further primary constraint(s). We derived the condition for this case in appendix \ref{appendix-detC}.
Another possibility is the presence of tertiary constraint as well as secondary constraint from our primary constraint. Yet another possibility is that our primary constraint and the corresponding secondary constraint become first class, that is the system possesses a gauge symmetry. The theory investigated in appendix \ref{App:EMtheory} will be
a concrete example of this case.
In order to clarify the general massless vector-tensor theories, we need to 
perform Hamiltonian analysis and we will address these issue in future.

Another possible direction will be to increase the number of derivatives in the Lagrangian.
In the present paper, for simplicity, we have only considered vector field theories, 
which contain up to two derivatives with respect to $g_{\mu \nu}$ and $A_\mu$. 
For example, one can consider a theory of the vector field with up to {\it three} derivatives with respect to $g_{\mu \nu}$ and $A_\mu$,
which corresponds to quintic-type theories in the GP or beyond GP theories.

In this paper, we have imposed the degeneracy of Lagrangian to eliminate the would-be Ostrogradski mode which can be carried by $A_*$. 
Even if Lagrangian contains an independent kinetic term of $\As$, that is, the theory is {\it non-degenerate}, the coefficient of the kinetic term of $\As$ can be possibly tuned as a positive value as well as other dynamical modes. This means that all of four propagating degrees of freedom have a proper sign of the kinetic terms. Of course, this does not guarantee the healthiness of the whole theory and we need to investigate the total Hamiltonian since its dangerous nature might appear in the form of tachyon and/or gradient instabilities. Concrete analysis of this type of theories, including the St\"uckelberg analysis and/or the Hamiltonian analysis will be also deferred to a future study.

\acknowledgments 
A.N. would like to thank Antonio De Felice and Shinji Mukohyama for fruitful discussions.
A.N. is grateful to Max-Planck-Institut f\"ur Astrophysik (MPA),
 Arnold Sommerfeld Center for Theoretical Physics (ASC),
 Academy of Sciences of the Czech Republic,  
 Laboratoire Astroparticule et Cosmologie (APC) and Hirosaki University for warm hospitality
 where this work was advanced and also the Yukawa Institute for Theoretical Physics at Kyoto University since discussions during the YITP workshop YITP-X-16-03 on "New perspective on theory and observation of large-scale structure" were useful for this work.
R.K. is supported by the Grant-in-Aid for Japan Society for the Promotion of Science (JSPS) Grant-in-Aid for Scientific Research Nos. 25287054. 
D.Y. is supported by the JSPS Research Fellowship for Young Scientists No. 2611495.
The work of A.N. is supported in part by the JSPS Research Fellowship for Young Scientists No. 263409
 and JSPS Grant-in-Aid for Scientific Research No. 16H01092.

\newpage

\appendix

\section{Details of transformations}
\label{appendix-trf.}
In this appendix, we study the detail
of metric transformations and vector field redefinition.
We first investigate the transformations of the general action (\ref{action}) in appendix \ref{appendix-gen}, and then	we focus on the transformations of the particular theories: Einstein-Maxwell theory in \ref{App:EMtheory} and the GP theory in \ref{App:GP}.

\subsection{Conformal and disformal metric transformations and vector field redefinition}
\label{appendix-gen}
Let us introduce the following transformations, 
\begin{align}
\bar{g}_{\mu\nu} &= \Omega(Y)g_{\mu\nu} + \Gamma(Y) A_{\mu} A_{\nu},\label{ghat=}\\
\bar{A}_{\mu} &= \Upsilon(Y) A_{\mu},\label{Ahat=}
\end{align}   
where $\Omega$, $\Gamma$ and $\Upsilon$ 
respectively represents conformal, disformal, and rescaling factors,
which are functions of $Y = A_\mu A^\mu$.
The transformations (\ref{ghat=}) and (\ref{Ahat=}) will be
the most general transformations which are constructed from $A_{\mu}$ and $g_{\mu\nu}$ and respect  general covariance.
The important feature is that the theory (\ref{action}) is closed under these transformations, that is, the transformations (\ref{ghat=}) and (\ref{Ahat=}) just cause changes in arbitrary functions of the theory, as we will explicitly see below.

Suppose that the original action (\ref{action}) is given in the barred frame, in which the fields and arbitrary functions are denoted as ${\bar g}_{\mu\nu},~{\bar A}_\mu,~{\bar f},~{\bar \alpha}_i$, and so on.
We then find transformation rules of the inverse metric and
the contravariant component of the barred vector as
\begin{align}
\bar{g}^{\mu\nu} &= \frac{1}{\Omega} \left( g^{\mu\nu} - \frac{\Gamma}{\Omega + Y \Gamma}A^{\mu}A^{\nu}\right), \\
\bar{A}^\mu &= \bar{g}^{\mu\nu}\bar{A}_{\nu}
=\frac{\Upsilon}{\Omega + Y \Gamma} A^{\mu}.
\end{align}
The determinant of two metrics are related by
\begin{align}
\sqrt{-\bar{g}} =\sqrt{-g} 
\sqrt{\Omega^3 (\Omega + Y \Gamma)}.
\label{sqrtg}
\end{align}
It is useful to relate $Y$ and  $\bar{Y}$, 
\begin{align}
\bar{Y} = \bar{g}^{\mu\nu}\bar{A}_{\mu}\bar{A}_{\nu} = \frac{Y \Upsilon^2 }{\Omega + Y \Gamma}.
\end{align}
Then, the barred covariant derivative of the barred vector 
is related to the unbarred covariant derivative through
\begin{align}
\bar{\nabla}_{\mu}\bar{A}_{\nu} = \nabla_{\mu} \bar{A}_{\nu} - {B}^{\rho}{}_{\mu\nu} \bar{A}_{\rho},
\end{align}
where we introduced ${B}^{\mu}{}_{\nu\rho} $, which is defined as
\begin{align}
{B}^{\mu}{}_{\nu\rho} = \frac{1}{2}\bar{g}^{\mu\sigma}\left( \nabla_{\nu}\bar{g}_{\sigma\rho}
+\nabla_{\rho}\bar{g}_{\sigma\nu}-\nabla_{\sigma}\bar{g}_{\nu\rho}\right).
\end{align}
The barred Riemann tensor can be expressed as
\begin{align}
\bar{R}_{\mu\nu\rho}{}^{\sigma} = R_{\mu\nu\rho}{}^{\sigma} -2 \nabla_{[\mu} {B}^{\sigma}{}_{\nu]\rho}+ 2{B}^{\lambda}{}_{\rho[\mu}{B}^{\sigma}{}_{\nu]\lambda},
\end{align}
where we use the following convention of the Riemann tensor:  $(\nabla_{\mu}\nabla_{\nu}-\nabla_{\nu}\nabla_{\mu}) V_{\rho} = R_{\mu\nu\rho}{}^{\sigma}V_\sigma$.

Plugging the above expressions into the action, we can 
express the arbitrary functions $f,\alpha_i$ in terms of the barred functions.
Since $\bar{R}$ contains the second derivative of $A_{\mu}$, we need to integrate $\bar{f} \bar{R} $ term of the action by parts in order to write the resultant action in the form of (\ref{action}). 
For the future reference, we write
the derivative of the function $\bar{f}$ with respect to $Y$ and $\bar{Y}$,
\begin{align}
\bar{f}_Y &= \frac{d\bar{Y}}{d Y}\frac{d \bar{f}}{d \bar{Y}} \notag\\
&= \left( \frac{\Upsilon^2 + 2 Y \Upsilon \Upsilon_Y}{\Omega + Y \Gamma} + \frac{Y \Upsilon^2(\Omega_Y + \Gamma + Y \Gamma_Y)}{(\Omega + Y \Gamma)^2}\right)\bar{f}_{\bar{Y}}.
\end{align}
We do not here consider the transformations of $\alpha_9, G_2$ and $G_3$ because these terms do not appear in the degeneracy condition.
For simplicity, we show the individual result of the conformal transformation, disformal transformation, and vector field redefinition.

\subsubsection*{Conformal Transformation : $(\Omega,\Gamma,\Upsilon) = (\Omega(Y), 0, 1)$}
\begin{subequations}
	\begin{align}
	f&=\Omega  \bar{f},\\
	\alpha _1&=\bar{\alpha }_1,\\
	\alpha _2&=\bar{\alpha }_2,\\
	\alpha _3& = 2 \frac{\Omega _Y}{\Omega }\left(\bar{\alpha }_1+2\bar{\alpha }_2\right)+\left(\frac{Y}{\Omega}\right)_Y\bar{\alpha }_3,\\
	\alpha _4&= 6 \Omega _Y \left(2 \left(\frac{Y}{\Omega}\right)_Y \bar{f}_{\bar{Y}}+\frac{\Omega _Y}{\Omega } \bar{f}\right)+2\left(\left(\frac{Y}{\Omega}\right)_Y\Omega_Y - \frac{\Omega_Y}{\Omega}\right) \bar{\alpha }_1+\left(\left(\frac{Y}{\Omega}\right)_Y\right)^2 \Omega \bar{\alpha }_4,\\
	\alpha _5&=2 \frac{\Omega _Y^2}{\Omega ^2}\left(\bar{\alpha }_1+2 \bar{\alpha }_2\right)+2\left(\frac{Y}{\Omega}\right)_Y \frac{\Omega _Y}{\Omega}\bar{\alpha }_3+\left(\left(\frac{Y}{\Omega}\right)_Y\right)^2\bar{\alpha }_5,\\
	\alpha _6&=\bar{\alpha }_6,\\
	\alpha _7&=6 \Omega _Y \left( 2 \left(\frac{Y}{\Omega}\right)_{Y} \bar{f}_{\bar{Y}}+  \frac{\Omega _Y}{\Omega} \bar{f}\right)
	+\frac{Y  \Omega _Y}{\Omega ^2} \left(
	2  \Omega _Y \bar{\alpha }_1+Y \frac{ \Omega _Y}{\Omega } \bar{\alpha }_4 + \bar{\alpha }_8 \right) + \frac{1}{\Omega }\bar{\alpha }_7 ,\\
	\alpha _8&=-12 \Omega _Y \left(2 \left(\frac{Y}{\Omega}\right)_{Y} \bar{f}_{\bar{Y}}+  \frac{\Omega _Y}{\Omega } \bar{f}\right)
	+\left( \frac{Y}{\Omega} \right)_Y \left(
	4 \Omega _Y \bar{\alpha }_1
	+2 Y  \frac{\Omega _Y}{\Omega} \bar{\alpha }_4 
	+ \bar{\alpha }_8\right).
	\end{align}
\end{subequations}

\subsubsection*{Disformal Transformation : $(\Omega,\Gamma,\Upsilon) = (1, \Gamma(Y), 1)$}
\begin{subequations}
	\begin{align}
	f=& \frac{\bar{f}}{J},\\
	\alpha _1=& \Gamma J \bar{f}+J^3 \bar{\alpha }_1,\\
	\alpha _2=&-\Gamma J \bar{f}+J^3 \bar{\alpha }_2,\\
	\alpha _3=&- 2 J\left( \Gamma _Y \bar{f} +2\Gamma (J^2 Y)_Y\bar{f}_{\bar{Y}}\right) + 4 J^2 J_Y  \bar{\alpha }_2+(J^2 Y)_Y  J^3 \bar{\alpha }_3,\\
	\alpha _4=& 2 J \left( \Gamma _Y \bar{f} - 4 J_Y (J^2 Y)_Y\bar{f}_{\bar{Y}}\right)
	+2 J \Bigl(2 J J_Y - Y \Gamma_Y  (J^2 Y)_Y \Bigr) \bar{\alpha }_1
	+\frac{((J^2 Y)_{Y})^2}{J} \bar{\alpha }_4 ,\\
	\alpha _5=&-4 \Gamma _Y  J (J^2 Y)_Y  \bar{f}_{\bar{Y}}+2 J \Bigl(2 (J_Y)^2 +  \Gamma_Y (J^2 Y)_Y \Bigr) \bar{\alpha }_1+ 4 J (J_Y)^2 \bar{\alpha }_2 \notag\\& +2 J^2(J^2 Y)_Y J_Y\bar{\alpha }_3+((J^2 Y)_Y)^2 J \left(-\Gamma   \bar{\alpha }_4+ J^2 \bar{\alpha }_5\right),\\
	\alpha _6=&-\frac{\Gamma}{J}  \bar{f} +\frac{1}{J}\bar{\alpha }_6 ,
	\displaybreak[1]\\
	\alpha _7=&2 J \left(\left(\Gamma ^2-\Gamma _Y\right)\bar{f} - 4 \frac{J_Y}{J^3} (J^2 Y)_Y  \bar{f}_{\bar{Y}}\right)+ \frac{8Y}{J} (J_Y)^2\bar{\alpha }_1+4 Y^2 J (J_Y)^2 \bar{\alpha }_4
	\notag\\&
	-2 \Gamma J  \bar{\alpha }_6 +J^3 \bar{\alpha }_7-2 Y J^2 J_Y \bar{\alpha }_8,\\
	\alpha _8=&  \frac{( J^2 Y)_Y}{J^2} \left(4 J_Y (4\bar{f}_{\bar{Y}} 
	-2 \bar{\alpha }_1- Y J^2 \bar{\alpha }_4)+J^3 \bar{\alpha }_8\right),
	\end{align}
	where we introduced $ J = 1/\sqrt{1+ \Gamma Y}$.
\end{subequations}

\subsubsection*{{Vector Field Redefinition} : $(\Omega, \Gamma, U) = (1, 0, \Upsilon(Y))$}
\begin{subequations}
	\begin{align}
	f=&\bar{f},\\
	\alpha _1=&\Upsilon ^2 \bar{\alpha }_1, \\
	\alpha _2=&\Upsilon ^2 \bar{\alpha }_2, \\
	\alpha _3=&4 \Upsilon  \Upsilon _Y \bar{\alpha }_2+\Upsilon ^3 \left(\Upsilon +2 Y \Upsilon _Y\right) \bar{\alpha }_3,\\
	\alpha _4=&2  \Upsilon _Y \left(2 \Upsilon +Y \Upsilon _Y\right)\bar{\alpha }_1+\Upsilon ^2  \left(\Upsilon +Y \Upsilon _Y\right){}^2\bar{\alpha }_4 +2 Y  \Upsilon _Y^2\bar{\alpha }_6 \notag\\
	& \qquad +Y^2 \Upsilon ^2  \Upsilon _Y^2 \bar{\alpha }_7 - Y \Upsilon ^2   \Upsilon _Y \left(\Upsilon +Y \Upsilon _Y\right)\bar{\alpha }_8, \\
	\alpha _5=&2 \Upsilon _Y^2 \bar{\alpha }_1+4 \Upsilon _Y^2 \bar{\alpha }_2+2 \Upsilon ^2  \Upsilon _Y \left(\Upsilon +2 Y \Upsilon _Y\right)\bar{\alpha }_3 +\Upsilon ^2  \Upsilon _Y \left(2 \Upsilon +3 Y \Upsilon _Y\right)\bar{\alpha }_4 \notag\\
	&+\Upsilon ^4  \left(\Upsilon +2 Y \Upsilon _Y\right){}^2\bar{\alpha }_5-2 \Upsilon _Y^2  \bar{\alpha }_6-Y \Upsilon ^2  \Upsilon _Y^2  \bar{\alpha }_7+\Upsilon ^2 \Upsilon _Y \left(\Upsilon +Y \Upsilon _Y\right)  \bar{\alpha }_8, \\
	\alpha _6=&\Upsilon ^2 \bar{\alpha }_6,\\
	\alpha _7=&2 Y  \Upsilon _Y^2 \bar{\alpha }_1 +  Y^2 \Upsilon ^2 \Upsilon _Y^2  \bar{\alpha }_4+2 \Upsilon _Y \left(2 \Upsilon +Y \Upsilon _Y\right) \bar{\alpha }_6 \notag\\
	& \qquad
	+\Upsilon ^2 \left(\Upsilon +Y \Upsilon _Y\right){}^2 \bar{\alpha }_7-Y \Upsilon ^2  \Upsilon _Y \left(\Upsilon +Y \Upsilon _Y\right) \bar{\alpha }_8,\\
	\alpha _8=&-4 \Upsilon _Y \left(\Upsilon +Y \Upsilon _Y\right) \bar{\alpha }_1-2 Y \Upsilon ^2  \Upsilon _Y \left(\Upsilon +Y \Upsilon _Y\right)\bar{\alpha }_4
	-4  \Upsilon _Y \left(\Upsilon +Y \Upsilon _Y\right)\bar{\alpha }_6   
	\notag\\&
	-2 Y \Upsilon ^2  \Upsilon _Y \left(\Upsilon +Y \Upsilon _Y\right)\bar{\alpha }_7 +\Upsilon ^2  \left(\Upsilon ^2+2 Y^2 \Upsilon _Y^2+2 Y \Upsilon  \Upsilon _Y\right)\bar{\alpha }_8.
	\end{align}
\end{subequations}

\subsection{Transformation of the Einstein-Maxwell theory}
\label{App:EMtheory}
Let us investigate the metric transformations (\ref{ghat=}) and the vector field redefinition (\ref{Ahat=}) of the Einstein-Maxwell System
\begin{align}
{\cal L} = \sqrt{-\bar{g}}\left(\frac{1}{2} \bar{R} - \frac{1}{4}\bar{F}_{\mu\nu}\bar{F}^{\mu\nu} \right).
\end{align}
This system can be found when the arbitrary functions are given by
\begin{align}
\bar{f} =& \frac{1}{2},~\bar{\alpha}_6 = -1,~\bar{\alpha}_1 = \bar{\alpha_2} = \bar{\alpha}_3=\bar{\alpha_4}=\bar{\alpha_5} = \bar{\alpha_7} = \bar{\alpha_8} = 0.
\end{align}
The arbitrary functions in the new frame are given by
\begin{align}
\alpha _1&=\alpha _2=\alpha _3=\alpha_5=0 \,, \quad
2 \alpha _4 = 2 \alpha_7 = - \alpha_8 = \frac{6 \Omega _Y^2}{\Omega } \,, \quad
\alpha _6 =-1 \,, \quad
f=\frac{\Omega }{2},
\end{align}
under the conformal transformation, $(\Omega\,, \Gamma \,, \Upsilon) = (\Omega(Y) \,, 0 \,, 1)$, and 
\begin{align}
\alpha _1&=-\alpha_2=\frac{\Gamma }{2 \sqrt{\Gamma  Y+1}}, \quad
\alpha _3=-\alpha_4=-\frac{\Gamma _Y}{\sqrt{\Gamma  Y+1}}, \quad
\alpha _5= \alpha_8=0, \notag\\
\alpha _6&=-\frac{1}{2} (\Gamma +2) \sqrt{\Gamma  Y+1}, \quad
\alpha _7=\frac{\Gamma ^2+2 \Gamma -\Gamma _Y}{\sqrt{\Gamma  Y+1}}, \quad
f=\frac{1}{2} \sqrt{\Gamma  Y+1},
\end{align}
under the disformal transformation, $(\Omega\,, \Gamma \,, \Upsilon) = (1 \,, \Gamma(Y) \,, 1)$, and 
\begin{align}
\alpha _1&=
\alpha _2=
\alpha _3=0, \quad
\alpha _4=-2 Y \Upsilon _Y^2, \quad
\alpha _5=2 \Upsilon _Y^2, \quad
\alpha _6=-\Upsilon ^2, \notag\\
\alpha _7&=-2 \Upsilon _Y \left(2 \Upsilon +Y \Upsilon _Y\right), \quad
\alpha _8=4 \Upsilon _Y \left(\Upsilon +Y \Upsilon _Y\right), \quad
f=\frac{1}{2},
\end{align}
under the vector field redefinition, $(\Omega\,, \Gamma \,, \Upsilon) = (1 \,, 0 \,, \Upsilon(Y))$.

The simplest but interesting example is the case with
$(\Omega\,, \Gamma \,, \Upsilon) = (1 \,, {-2} \,, 1)$
where the disformal factor $\Gamma = -2$ is chosen to maximally simplify the resultant action.
The resultant action is given by
\begin{align}
{\cal L} &= \sqrt{{-g}} \left( \frac{1}{2} \sqrt{1 {- 2} Y} \, R - {\frac{1}{4\sqrt{1-2Y}}} (S_{\mu \nu}^2 - S^2) \right). 
\end{align}
Note that the transformed theory is nothing but a subclass of the GP theory.
Interestingly, $U(1)$ gauge invariance in this theory is not transparent due to the explicit dependence on $Y$.
However, as studied in \cite{Domenech:2015tca}, as long as a metric transformation is invertible,
the nature of theory, namely the set of constraints and the associated constraint algebra,
should be left unchanged. 
Based on this argument, it will be quite plausible that $U(1)$ symmetry in the original theory
still exists even in the transformed theory
as an extended gauge symmetry accompanied by the metric tensor.
However, in order to reveal a hidden constraint in this system, further detailed investigation,
that is Hamiltonian analysis of such a vector-tensor theory will be necessary.
We will defer this interesting topic in future work.

\subsection{Transformation of the Generalized Proca theory}
\label{App:GP}
Let us consider the GP theory in barred frame,
namely,
\begin{align}
\bar{f} = G_4{}(\bar{Y}), \quad
\bar{\alpha}_1 = - \bar{\alpha}_2 = 2 G_4{}_{, \bar{Y}}(\bar{Y}), \quad
\bar{\alpha}_3 = \bar{\alpha}_4 = \bar{\alpha}_5=\bar{\alpha}_8 = 0,
\end{align}
with arbitrary $\bar{\alpha}_6$ and $\bar{\alpha}_7$. 
We first take a look at arbitrary functions in a new frame under the disformal transformation and then show the results when all of the transformations are performed simultaneously.
It should be noted that $\alpha_1 + \alpha_2$ vanishes 
under any of the transformations.

\subsubsection*{Disformal Transformation of the Generalized Proca Theory}
\begin{subequations}
	\begin{align}
	f &= G_4 \sqrt{\Gamma  Y+1},\\
	\alpha _1 &= -\alpha_2 = \frac{2 G_4{}_{\bar{Y}}+\Gamma  G_4 (\Gamma  Y+1)}{(\Gamma  Y+1)^{3/2}},\\
	\alpha _3 &= -\alpha_4 =-\frac{2 \Gamma _Y \left(G_4 (\Gamma  Y+1)-2 Y G_4{}_{\bar{Y}}\right)}{(\Gamma  Y+1)^{3/2}},\\
	\alpha _6 &= \bar{\alpha }_6 \sqrt{\Gamma  Y+1}-\Gamma  G_4 \sqrt{\Gamma  Y+1},\\
	\alpha _7 &= \frac{2 \left(2 \left(\Gamma +Y \Gamma _Y\right) G_4{}_{\bar{Y}}+G_4 (\Gamma  Y+1) \left(\Gamma ^2-\Gamma _Y\right)\right)}{(\Gamma  Y+1)^{3/2}}-\frac{2 \Gamma  \bar{\alpha }_6}{\sqrt{\Gamma  Y+1}}+\frac{\bar{\alpha }_7}{(\Gamma  Y+1)^{3/2}},\\
	\alpha _5 &= \alpha_8=0.
	\end{align}
\end{subequations}
Since
\begin{align}
f_Y =& \frac{G_4 (\Gamma  Y+1) \left(\Gamma +Y \Gamma _Y\right)-2 \left(Y^2 \Gamma _Y-1\right)
	G_4{}_{\bar{Y}}}{2 (\Gamma  Y+1)^{3/2}},
\end{align}
one can show that
\begin{align}
\alpha_1 = 2 f_Y + \frac{Y}{2} \alpha_3.
\label{disformalGP}
\end{align}
The above parameter choice (\ref{disformalGP}) is exactly the same as the condition of the beyond GP theory (\ref{caseB}).
Thus, the resultant theory belongs to the beyond GP theory. 
Note that the transformed theory with $\Gamma={\rm const.}$ is just the GP theory itself since $\alpha_3=\alpha_4=0$.

\subsubsection*{Transformations of the Generalized Proca Theory}
We now perform all kinds of transformations to the GP theory simultaneously.
The result is
\begin{subequations}\label{TGP}
	\begin{align}
	f &=G_4 \sqrt{\Omega(\Gamma  Y+\Omega) },
	\label{Tf}
	\\
	\alpha _1 &= - \alpha_2 =\frac{G_4 \Gamma\sqrt{\Omega }}{\sqrt{\Gamma  Y+\Omega }}+\frac{2 G_{4 \bar{Y}}\Upsilon ^2 \Omega ^{3/2}}{(\Gamma  Y+\Omega )^{3/2}},\label{Talpha_2}\\
	\alpha _3 &=-\frac{2 G_4 \left(\Omega  \Gamma _Y+\Gamma  \Omega _Y\right)}{\sqrt{\Omega }
		\sqrt{\Gamma  Y+\Omega }}-\frac{4 \Upsilon  \sqrt{\Omega } G_{4 \bar{Y}} \(2 \Upsilon _Y (\Gamma  Y+\Omega
		)+\Upsilon  \left(\Omega _Y-Y \Gamma _Y\right)\)}{(\Gamma  Y+\Omega
		)^{3/2}},\label{Talpha_3}\\
	\alpha _4 &=\frac{\bar{\alpha }_7 \Upsilon ^2 Y^2 \sqrt{\Omega }
		\Upsilon _Y^2}{(\Gamma  Y+\Omega )^{3/2}}+\frac{2
		\bar{\alpha }_6 Y \sqrt{\Omega } \Upsilon
		_Y^2}{\sqrt{\Gamma  Y+\Omega }}+\frac{2 G_4 \(\Omega  \Gamma _Y \left(2 Y \Omega
		_Y+\Omega \right)+\Omega _Y \bigl(\Gamma  \Omega
		+\Omega _Y (\Gamma  Y+3 \Omega
		)\bigr)\)}{\Omega ^{3/2} \sqrt{\Gamma 
			Y+\Omega }}\nonumber\\
	&~~~+4 G_{4 \bar{Y}} \[\frac{\Upsilon _Y \bigl(Y \Omega 
		\Upsilon _Y+2 \Upsilon  \left(2 Y \Omega _Y+\Omega
		\right)\bigl)}{\sqrt{\Omega } \sqrt{\Gamma 
			Y+\Omega }}-\frac{\Upsilon ^2 \left(Y \Gamma _Y \left(2 Y \Omega
		_Y+\Omega \right)+\Omega _Y \left(2 Y \Omega
		_Y-\Omega \right)\right)}{\sqrt{\Omega } (\Gamma 
		Y+\Omega )^{3/2}}\]
	,\label{Talpha_4}\\
	\alpha _5 &=-\frac{\bar{\alpha }_7 \Upsilon ^2 Y \sqrt{\Omega
		} \Upsilon _Y^2}{(\Gamma  Y+\Omega )^{3/2}}-\frac{2
		\bar{\alpha }_6 \sqrt{\Omega } \Upsilon
		_Y^2}{\sqrt{\Gamma  Y+\Omega }}-\frac{2 G_4 \Omega _Y
		\left(2 \Omega  \Gamma _Y+\Gamma  \Omega
		_Y\right)}{\Omega ^{3/2} \sqrt{\Gamma  Y+\Omega
		}}\nonumber\\
		&~~~~-\frac{4 G_{4 \bar{Y}} \left(\Upsilon ^2 \Omega _Y
			\left(\Omega _Y-2 Y \Gamma _Y\right)+\Gamma  Y
			\Upsilon _Y \left(\Omega  \Upsilon _Y+4 \Upsilon 
			\Omega _Y\right)+\Omega ^2 \Upsilon _Y^2+4 \Upsilon 
			\Omega  \Upsilon _Y \Omega _Y\right)}{\sqrt{\Omega }
			(\Gamma  Y+\Omega )^{3/2}},\label{Talpha_5}\\
		\alpha _6 &=\frac{\bar{\alpha }_6 \Upsilon ^2 \sqrt{\Gamma  Y+\Omega }}{\sqrt{\Omega }}-\frac{\Gamma
			G_4 \sqrt{\Gamma  Y+\Omega }}{\sqrt{\Omega }},\label{Talpha_6}\\
		\alpha _7 &=\frac{\bar{\alpha
			}_7 \Upsilon ^2 \sqrt{\Omega } \left(\Upsilon +Y
			\Upsilon _Y\right){}^2}{(\Gamma  Y+\Omega
			)^{3/2}}+\frac{\bar{\alpha }_6 \left(2 \Omega 
			\Upsilon _Y \left(2 \Upsilon +Y \Upsilon _Y\right)-2
			\Gamma  \Upsilon ^2\right)}{\sqrt{\Omega }
			\sqrt{\Gamma  Y+\Omega }}\nonumber\\
		&~~~~+\frac{2 G_4 \left(\Gamma ^2 \Omega +\Gamma 
			\Omega _Y \left(Y \Omega _Y+3 \Omega \right)+\Omega 
			\left(\Omega _Y \left(2 Y \Gamma _Y+3 \Omega
			_Y\right)-\Omega  \Gamma _Y\right)\right)}{\Omega
			^{3/2} \sqrt{\Gamma  Y+\Omega }}\nonumber\\
		&~~~~+\frac{1}{\sqrt{\Omega }(\Gamma  Y+\Omega )^{3/2}}
		\[4 G_{4 \bar{Y}} \(\Upsilon ^2 Y \Gamma _Y
		\left(\Omega -2 Y \Omega _Y\right)+\Upsilon ^2
		\left(\Gamma  \Omega -2 Y \Omega _Y^2+3 \Omega 
		\Omega _Y\right)\nonumber\\
		&~~~~~~~~~~~~~~~~~~~~~~~~~~~~~~
		+4 \Upsilon  Y \Upsilon _Y \Omega_Y (\Gamma  Y+\Omega )+Y \Omega  \Upsilon _Y^2	(\Gamma  Y+\Omega )\)\]
		,\label{Talpha_7}\displaybreak[1]\\
		\alpha _8 &=-\frac{4 G_4 \Omega _Y \left(\Omega 
			\left(2 Y \Gamma _Y+3 \Omega _Y\right)+\Gamma 
			\left(Y \Omega _Y+2 \Omega \right)\right)}{\Omega
			^{3/2} \sqrt{\Gamma  Y+\Omega }}\nonumber\\
		&~~~~-\frac{8 G_{4 \bar{Y}} \left(-2 \Upsilon ^2 \Omega _Y \left(Y \left(Y \Gamma
			_Y+\Omega _Y\right)-\Omega \right)+\Upsilon 
			\Upsilon _Y \left(4 Y \Omega _Y+\Omega \right)
			(\Gamma  Y+\Omega )+Y \Omega  \Upsilon _Y^2
			(\Gamma  Y+\Omega )\right)}{\sqrt{\Omega } (\Gamma  Y+\Omega
			)^{3/2}}\nonumber\\
		&~~~~-\frac{2
			\bar{\alpha }_7 \Upsilon ^2 Y \sqrt{\Omega }
			\Upsilon _Y \left(\Upsilon +Y \Upsilon
			_Y\right)}{(\Gamma  Y+\Omega )^{3/2}}-\frac{4
			\bar{\alpha }_6 \sqrt{\Omega } \Upsilon _Y
			\left(\Upsilon +Y \Upsilon
			_Y\right)}{\sqrt{\Gamma  Y+\Omega }}\label{Talpha_8}.
		\end{align}
	\end{subequations}

	\subsection{Transformation to Simple Frame}
	\label{appendix-simple}
In this appendix, we see that it is possible to move to a simple frame, for example, the frame with $f = 1$ starting from a non-trivial $f$ through transformations.
	\subsubsection{$f=1$ frame}
	Let us consider a conformal transformation. The parameter $f$ transforms as  
	\begin{align}
	f = \bar{f}\left(\frac{Y}{\Omega(Y)}\right) \Omega(Y). 
	\end{align}
	Then,  when 
	$\bar{f} \neq c \bar{Y}$, where $c$ is a constant, we can move to $f =1 $ frame by a conformal transformation with $\Omega$ which satisfies
	\begin{align}
	1 = \bar{f}\left(\frac{Y}{\Omega(Y)}\right) \Omega(Y).\label{tof=1}
	\end{align}
	Even in the case of $\bar{f} = c\bar{Y}$, we can move to $f =1$ frame by a disformal transformation with
	\begin{align}
	\Gamma = -\frac{1}{Y} + c^2Y.
	\end{align} 
	Actually, $f$ is transformed as
	\begin{align}
	f = \bar{f}\left(\frac{Y}{1+ Y \Gamma}\right) \sqrt{1 + Y \Gamma} = \frac{c Y}{\sqrt{1 + Y \Gamma}} = 1.
	\end{align}
	Therefore, we can set $f=1$
	 without loss of generality.

	\subsubsection{$f=1$ and $\beta=0$ frame}
	Let us consider further transformation from $\bar{f} = 1$ frame. We  focus on the combination
	\begin{align}
	\bar{\beta} =-2  \bar{\alpha}_6 - \bar{Y} \bar{\alpha}_7\,\neq 0.
	\end{align}
	After a vector field redefinitions, $f$ {and} $\beta$ become
	\begin{align}
	f =& 1,\\
	\beta =& \Upsilon ^2 \bar{\beta }+ 2 Y \Upsilon \Upsilon _Y\bar{\beta }+ Y^2
	\Upsilon _Y\Upsilon ^3  \bar{\alpha }_8+Y^2 \Upsilon _Y^2 \left(\bar{\beta}-2 \bar{\alpha }_1+\Upsilon ^2 Y \left(\bar{\alpha }_8-\bar{\alpha }_4\right)\right).
	\label{f=1,L=}
	\end{align}
	Although this first-order differential equation is quadratic in $\Upsilon_Y$, one can easily show that the discriminant can be positive by choosing an appropriate initial condition of $\Upsilon$. 
	Therefore, there exists a real non-zero solution $\Upsilon$ with $\beta=0$, and we can set $\beta = 0$ by a vector field redefinition without loss of generality.

\section{Determinants in Case A and B}
\label{appendix-detC}
In this appendix, we summarize the determinant of $\M_1$ and $y_1$ in each case.
Because of its complexity, 
the arbitrary functions are set to be $f=1$ and $\beta=0$,
which can be always set after transformations as studied in appendix \ref{appendix-simple}. \\
~\\
{\bf  \underline{Case~A1} :}
\ba
&&
f=1, \quad \alpha_1=-\alpha_2={1\over Y}, \quad \alpha_3={2\over Y^2}, \quad \beta=0,\notag\\
&& \det \M_1=\frac{(Y+\As^2)^2}{4Y^4}, \notag\\
&&y_1=-{1\over 16 Y^5}(Y+\As^2) \(Y^2+(1+2Y)\As^2+\As^4\)\notag\\
&&~~~~~~~\times
\[
Y\(16+8Y^2(\alpha_4-\alpha_8)+Y^4\alpha_8^2\)\notag\\
&&~~~~~~~~~~~
+\(4Y^5\alpha_5(\alpha_4+\alpha_8)-32-8Y^2\alpha_4-24Y^3\alpha_5+Y^4(2\alpha_4+\alpha_8)^2\)\As^2\].
\ea
The scalar sector further degenerates if 
$\alpha_4=-(4-Y^2\alpha_8)^2/8Y^2$ and $\alpha_5=-\alpha_8(8-Y^2\alpha_8)/8Y$, or 
$\alpha_4=-2/Y^2$ and $\alpha_8=8/Y^2$.\\
~\\
{\bf  \underline{Case~A2} :}
\ba
&&
f=1, \quad \alpha_1=-\alpha_2={1\over Y}, \quad \alpha_4={6\over Y^2}-\alpha_8,\notag\\
&& \det \M_1=\frac{(Y+\As^2)^2}{4Y^4}, \notag\\
&&y_1=-{1 \over 16 Y^5}
\[
Y^4 (\alpha _8 Y^2-8){}^2
+ Y^2 \(4 (\alpha _3^2+\alpha _8 \alpha _3+\alpha
 _8^2) Y^7+(12 \alpha _3^2-4 \alpha _8
 \alpha _3+\alpha _8^2) Y^6-24 (2 \alpha
 _3+3 \alpha _8) Y^5\notag\\
&&~~~~~~~~~~~~~~~~~~ -8 (2 \alpha _3+\alpha
 _8) Y^4+336 Y^3+48 Y^2\){A}_*^2
+2Y^2\(
(4 \alpha _3^2+6 \alpha _8 \alpha _3+3 \alpha
_8^2) Y^5+(\alpha _8^2-4 \alpha
_3^2) Y^4\notag\\
&&~~~~~~~~~~~~~~~~~~
-(64 \alpha _3+60 \alpha
_8) Y^3+16 (\alpha _3-\alpha _8)
Y^2+304 Y+48
\){A}_*^4
\notag\\
&&~~~~~~~~~~~~~~~~~~+
 \((2 \alpha _3+\alpha _8)
Y^2-12\) \(Y \bigl(4 (\alpha _3+\alpha
_8) Y^2+(2 \alpha _3+\alpha _8)
Y-40\bigr)-12\){A}_*^6
\notag\\
&&~~~~~~~~~~~~~~~~~~
+\((2 \alpha _3+\alpha _8)
Y^2-12\){}^2 {A}_*^8 
\].
\ea
The scalar sector further degenerates when $\alpha_3=2/Y^2$ and $\alpha_8=8/Y^2$. \\
~\\
{\bf  \underline{Case~A3} :}
\begin{align}
&
f=1, \quad \alpha_1=-\alpha_2=-{1 \over 2} (\alpha_4+\alpha_8)Y, \notag\\
& \alpha_3=-{1\over 4}(4\alpha_4+2\alpha_8)+
{1\over 4}\alpha_8(\alpha_4+\alpha_8),
\notag\\
& \det \M_1=\frac{
	\(2\alpha_6+Y(Y\alpha_6-1)(\alpha_4+\alpha_8)\)^2
	(Y+\As^2)^2}{16Y^2},\notag\displaybreak[2]\\
&y_1=-{1\over 128}
\[2\alpha_8(Y^2\alpha_8-16)+\(3 \alpha _8^2 \left(\alpha _4+\alpha
_8\right) Y^3-64 \alpha _5-8 \alpha _8 \left(2 \alpha _4+\alpha
_8\right) Y\)\As^2\]\notag\\
&~~~~~~~\times \[4Y+ \(8 - 4 Y (\alpha_4 + \alpha_8)
 + Y^2(1+Y)^2 (\alpha_4+\alpha_8)^2 \) \As^2 +Y^2(\alpha_4+\alpha_8)^2 \As^4 \].
\end{align}
The scalar sector further degenerates when $\alpha_5=\alpha_8=0$, $\alpha_5=8(20+Y^2\alpha_4)/Y^3$ and $\alpha_8=16/Y^2$, or $\alpha_4=-\alpha_8=-\alpha_5/2Y=-16/Y^2$.
\\
~\\
{\bf  \underline{Case~A4} :}\\
As proved in section \ref{trfA4}, 
the case A4 can be mapped to the GP theory through the transformations (\ref{ghat=}) and (\ref{Ahat=}).
Therefore, it is not needed to explicitly show the results of the determinant of $\M_1$ and $y_1$ here since 
these properties will not change though transformations as long as a transformation is invertible. 
The determinant of $\M_1$ and $y_1$ are already shown in section \ref{sec:GP}.
If a theory cannot be obtained from an invertible transformation, one needs to directly check the determinant of ${\cal M}_1$ and $y_1$ in the case A4. As an example, in the case of (\ref{singularA4}), 
	\ba
	 &&\det \M_1={1\over 16}\[ 2(\alpha_6 +\alpha_2 \alpha_6-\alpha_2) Y
	 +\(6+4\alpha_2-\alpha_7 Y- \alpha_2 \alpha_7 Y\) \As^2\]^2, \nonumber\\
	&&y_1={1\over 2} (1+\alpha_2)^2 (6+\beta)(3+\As^2)\As^2.
	\ea

~\\
{\bf  \underline{Case~B1} :}
\begin{align}
&f=1, \quad \beta=0, \quad 
\alpha_4=-{2\alpha_1 \over Y}, \quad
\alpha_5=\frac{(2\alpha_1-Y\alpha_3)^2}{4Y^2 (\alpha_1+\alpha_2)}, \quad \alpha_8=0,
\notag\\
& \det \M_1= \frac{(\alpha_1+\alpha_6-Y\alpha_1\alpha_6)^2(Y+\As^2)^2}{4Y^2},\notag\displaybreak[1]\\
&y_1={1\over 2Y^2(\alpha_1+\alpha_2)}
\[
4Y^2(\alpha_1+\alpha_2)^2
+Y\(4\alpha_1\bigl(2\alpha_2(1+Y\alpha_2)+\alpha_1(3+2Y\alpha_2)\bigr)\notag\\
&~~~~~~~~~~~~~~~~~~~~~~~~~~~
+4Y\bigl(\alpha_1Y(\alpha_1+\alpha_2)-2\alpha_1-\alpha_2\bigr)\alpha_3+Y^2\alpha_3^2\)\As^2\notag\\
&~~~~~~~~~~~~~~~~~~~~~~~~~~~
+\(2Y^2\alpha_3^2-8Y\alpha_1\alpha_3+\alpha_1^2\bigl(8+Y(1+Y)(2\alpha_2+Y\alpha_3)^2\bigr)\)\As^4\notag\\
&~~~~~~~~~~~~~~~~~~~~~~~~~~~
+\alpha_1(2\alpha_2+Y\alpha_3)\(2\alpha_1(\alpha_2+Y\alpha_2-2)+Y(2+\alpha_1+Y\alpha_1)\alpha_3\)\As^6\notag\\
&~~~~~~~~~~~~~~~~~~~~~~~~~~~
+\alpha_1^2(2\alpha_2+Y\alpha_3)^2\As^8\].
\end{align}
In this case, $y_1$ is always non-zero for any free functions.\\
~\\
{\bf  \underline{Case~B2} :}
\ba
&&f=1, \quad \beta=0, \quad 
\alpha_1={1\over Y}, \quad \alpha_4=-{2\over Y^2},
\quad \alpha_8={8\over Y^2},
\notag\\
&& \det \M_1={(Y+\As^2)^2 \over 4Y^4}, \notag\\
&&y_1={1\over 4Y^5}
\[
8Y(1+Y\alpha_2)+\(4+8Y\alpha_2+4Y^2\alpha_3-Y^4\alpha_3^2+4Y^3(1+Y\alpha_2)\alpha_5\)\As^2\]\notag\\
&&~~~~~~~\times
\(Y^2(3+Y)+2(Y-1)Y\As^2+(1+2Y)\As^4+\As^6\).
\ea
The scalar sector further degenerate when $\alpha_2=-1/Y$ and $\alpha_3=2/Y^2$.
\\
~\\
{\bf  \underline{Case~B3} :}
\begin{align}
&f=1, \quad \beta=0, \quad 
\alpha_1={1\over Y}, \quad \alpha_4=-{2\over Y^2}+\alpha_8-{Y^2\alpha_8^2 \over 8},\notag\\
&\alpha_5=\frac{(2-Y^2\alpha_3)^2}{4Y^3(1+Y\alpha_2)}
-{\alpha_8 \over Y}+{Y\alpha_8^2 \over 8},\notag\\
& \det \M_1={(Y+\As^2)^2 \over 4Y^4}, \notag\displaybreak[1]\\
&y_1=\frac{1} {128 Y^3 (\alpha _2 Y+1)}({A}_*^2+Y)\notag\\
&~~~~~~~
\[
4 Y \left(\alpha _2 Y+1\right){}^2\(\alpha _8^2 (4 Y+3) Y^3-32 \alpha _8 Y^2+64\)\notag\\
&~~~~~~~~
+\biggl(512+\alpha_3^2 Y^4 \left(\alpha _8 Y^2-8\right){}^2
-4 \alpha _3 Y^2 \left(\alpha _8 Y^2-8\right) 
\(
20 \alpha _2^2 \alpha _8^2 Y^6+48 \alpha _2 \alpha _8^2
Y^5\notag\\
&~~~~~~~~
-8 \alpha _8 \left(16 \alpha _2^2+\left(\alpha
_2-4\right) \alpha _8\right) Y^4-\alpha _8
\left(\alpha _2 \left(64 \alpha _2+319\right)+8
\alpha _8\right) Y^3\notag\\
&~~~~~~~~
+\left(256 \alpha _2^2-65 \alpha
_8 \alpha _2-254 \alpha _8\right) Y^2+\left(512
\alpha _2-\alpha _8\right) Y-8
\)
\biggr)\As^2
\notag\\
&~~~~~~~~
+\(16 \alpha _2-\alpha _3 \alpha _8 Y^3+2 \left(4 \alpha
_3+\alpha _8\right) Y\)
\biggl(16 \left(\alpha _2-2\right)-2 \alpha _3 \alpha _8
Y^4+\left(4 \alpha _2-\alpha _3\right) \alpha _8
Y^3\notag\\
&~~~~~~~~+8 \left(2 \alpha _3+\alpha _8\right) Y^2+2
\left(4 \alpha _3+\alpha _8\right) Y
\biggr)\As^4
\notag\\
&~~~~~~~~+
\left(16 \alpha _2-\alpha _3 \alpha _8 Y^3+2 \left(4
\alpha _3+\alpha _8\right) Y\right){}^2\As^6
\].
\end{align}
In this case, $y_1$ is always non-zero for any free functions.\\
~\\
{\bf  \underline{Case~B4} :}
\begin{align}
&f=1, \quad \beta=0, \quad 
\alpha_4=-{16 \over Y^2}-{2\alpha_1 \over Y},\quad \alpha_8={16 \over Y^2},\notag\\
&\alpha_5=\frac{
	64+Y\[4\alpha_1(8-7Y\alpha_1)-64(-2+Y\alpha_1)\alpha_2+4Y(-4+3Y\alpha_1)\alpha_3+Y^3\alpha_3^2\]}{4Y^4(\alpha_1+\alpha_2)},
\notag\\
& \det  \M_1=\frac{(\alpha_1+\alpha_6-Y\alpha_1\alpha_6)^2(Y+\As^2)^2}{4Y^2}, \notag\displaybreak[1]\\
&y_1=\frac{1}{2 (\alpha_1+\alpha _2) Y^4}
\[12 \left(\alpha _1+\alpha _2\right){}^2 Y^3 (3 Y+4)
+\biggl(
\alpha _3^2 Y^5+\(
4 \alpha _1^2 (3 Y+2) Y^4\notag\\
&~~~~~~~~~~~~~~~~~~~~~~~~~~
+4 \alpha _1 Y^4 \bigl(\alpha_2 (3 Y+2)-2\bigr)-4 Y^3 \left(5 \alpha _2Y+4\right)\)\alpha_3\biggr)\As^2\notag\\
&~~~~~~~~~~~~~~~~~~~~~~~~~~
+\biggl(
\(\alpha _1^2 Y (Y+1)+2\)\alpha _3^2 Y^4 
-\(8 \alpha _1^3 (Y+1) Y^4-24 \alpha _1 \alpha _2 (Y+1)
Y^3\notag\\
&~~~~~~~~~~~~~~~~~~~~~~~~~~
+4 \alpha _1^2 Y^3 (Y+1) \left(5 \alpha _2
Y-2\right)+8 Y^2 \left(3 \alpha _2 Y+4\right)\) \alpha_3\notag\\
&~~~~~~~~~~~~~~~~~~~~~~~~~~
+4 \alpha _1^2 (2 \alpha _1+5 \alpha _2){}^2
Y^4+4 \alpha _1 \(4 \alpha _1^3+4 (5 \alpha
_2-2) \alpha _1^2\notag\\
&~~~~~~~~~~~~~~~~~~~~~~~~~~
+\alpha _2 (25 \alpha
_2-44) \alpha _1-60 \alpha _2^2\) Y^3+4
\((2-44 \alpha _2)
\alpha _1^2-8 \alpha _1^3\notag\\
&~~~~~~~~~~~~~~~~~~~~~~~~~~+20 (1-3 \alpha _2) \alpha _2
\alpha _1+52 \alpha _2^2\) Y^2+64 \alpha _2
(\alpha _1+2 \alpha _2+3) Y+128
\biggr)\As^4
\notag\\
&~~~~~~~~~~~~~~~~~~~~~~~~~~
+
\(8 \alpha _2+\alpha _1 Y \left(\alpha _3 Y-4 \alpha _1-10 \alpha_2\right)\)
\biggl(
2 \left(4 \alpha _2+\alpha _3 Y^2-8\right)-4 \alpha _1^2
Y (Y+1)\notag\\
&~~~~~~~~~~~~~~~~~~~~~~~~~~+\alpha _1 Y \left((Y+1) \left(\alpha _3 Y-10
\alpha _2\right)+4\right)
\biggr)\As^6
\notag\\
&~~~~~~~~~~~~~~~~~~~~~~~~~~
+\biggl(8 \alpha _2+\alpha _1 Y \left(\alpha _3 Y-4 \alpha _1-10
\alpha _2\right)\biggr)^2\As^8
\].
\end{align}
In this case, $y_1$ is always non-zero for any free functions.\\
~\\
{\bf  \underline{Case~B5} :}
\begin{align}
&f=1, \quad \beta=0, \quad 
\alpha_1=\frac{
	-8(2\alpha_2+Y\alpha_3)+Y(-4-4Y\alpha_2+Y^2\alpha_3)\alpha_8}{2Y^2\alpha_8}\notag\\
&\alpha_4={4(1+Y\alpha_2) \over Y^2}-\alpha_3+{8(2\alpha_2+Y\alpha_3) \over Y^3 \alpha_8}+\alpha_8-{Y^2\alpha_8^2 \over 8},\notag\\
&\alpha_5={-2+Y^2 \alpha_3 \over Y^3}-{4(2\alpha_2+Y\alpha_3) \over Y^4 \alpha_8}-{\alpha_8 \over Y}+{Y\alpha_8^2 \over 8}+{12(2\alpha_2+Y\alpha_3) \over Y^2(-8+Y^2\alpha_8)},
\notag\displaybreak[1]\\
& \det \M_1=\frac{1}{4096 \alpha_8^2 Y^6}
\[16 \alpha _6 Y^2 \(\alpha _8 Y (\alpha _3 Y^2-4\alpha _2 Y-6)-8 (2 \alpha _2+\alpha _3Y)\)\notag\\
&~~~~~~~~~~~~~~~~~~~~~~~~~~~~
+16 Y \(4\alpha _2 (\alpha _8 Y^2+4)-\alpha _3 \alpha _8 Y^3+4 (2\alpha _3+\alpha _8) Y\)\notag\\
&~~~~~~~~~~~~~~~~~~~~~~~~~~~~
+\biggl(
96 \alpha _8^2 Y^3-6 \alpha _8^3 Y^5+32 \alpha _8 Y
\left(2-3 \alpha _6 Y\right)\notag\\
&~~~~~~~~~~~~~~~~~~~~~~~~~~~~
+\(\alpha _8 Y^2 \left(\alpha _3 Y-4 \alpha _2\right)-8
\left(2 \alpha _2+\alpha _3 Y\right)\)\(\alpha _8^2 Y^4-16 \alpha _8 Y^2+16 \left(\alpha _6
Y-1\right)\)
\biggr)\As^2\]^2,\notag\\
~\displaybreak[1]\\
&y_1=-\frac{1}{128	\alpha _8 Y^4 (\alpha _8 Y^2-8)}
\[2 Y (\alpha _8 Y^2-8)\(2 \alpha _2(\alpha _8 Y^2+8)+8 \alpha _3 Y+4\alpha _8 Y-\alpha _3 \alpha _8 Y^3\)\notag\\
&~~~~~~~~~~~~~~~~~~~~~~~~~~~~~~~~~~~~~
+{A}_*^2\biggl(\alpha _8^2 Y^3 \(Y \bigl(\alpha_3^2 Y^3-2 \alpha _3 Y (\alpha _2Y+4)-8 \alpha _2 (\alpha _2Y+1)\bigr)+8\)\notag\\
&~~~~~~~~~~~~~~~~~~~~~~~~~~~~~~~~~~~~~
+8 \alpha _8 Y\(Y \bigl(-3 \alpha _3^2 Y^3+16 \alpha
_3 Y+12 \alpha _2 (\alpha _2Y+2)\bigr)-8\)\notag\\
&~~~~~~~~~~~~~~~~~~~~~~~~~~~~~~~~~~~~~+128 (2\alpha _2+\alpha _3 Y) (\alpha _3Y^2+2 \alpha _2 Y-1)\biggr)\]
\notag\\
&~~~~~~~~~~~~~~~~~~~~~~~~~~~~~~~~~~~~~
\times\[\alpha _8^2 (4 Y+3) Y^4-32 \alpha _8Y^3+{A}_*^2 \biggl(\alpha _8 Y^2\(\alpha _8 Y (5 Y+4)-16\)\notag\\
&~~~~~~~~~~~~~~~~~~~~~~~~~~~~~~~~~~~~~~~~~
+4 \alpha_8 {A}_*^2 Y\(\alpha _8 Y(Y+1)+\alpha _8 {A}_*^2Y+8\)+128\biggr)+64 Y\].
\end{align}
The scalar sector further degenerates when $\alpha_2=-1/Y$ and $\alpha_3=2/Y^2$ or $\alpha_2=(Y^2 \alpha_3 -8)/6Y$ and $\alpha_8=16 / Y^2$.
\\
~\\
{\bf  \underline{Case~B6} :}
\begin{align}
&f=1, \quad \beta=0, \quad 
\alpha_2={-2-Y\alpha_1 \over 3Y}, \quad \alpha_3={2(2+Y\alpha_1 ) \over 3Y^2},\quad
\alpha_4=-{2\alpha_1 \over Y}+\alpha_8-{Y^2\alpha_8^2 \over 8},\notag\\
&\alpha_5={2(-1+Y\alpha_1)\over 3Y^3}-{\alpha_8 \over Y}+{Y \alpha_8^2 \over 8},
\notag\\
& \det \M_1={1\over 1024Y^2}
\[16Y\(-\alpha_6+\alpha_1(-1+Y\alpha_6)\)\notag\\
&~~~~~~~~~~~~~~~~~~~~~~~~~
+\(-16\alpha_6+16\alpha_1(-1+Y\alpha_6)-16Y(-1+Y\alpha_1)\alpha_8+Y^3(-1+Y\alpha_1)\alpha_8^2\)\As^2
\]^2,\notag\displaybreak[1]\\
&y_1={1\over 48Y^3} (-1+Y\alpha_1)(Y+\As^2)\notag\\
&~~~~~~\times
\[64Y-32Y^3\alpha_8+Y^4(3+4Y)\alpha_8^2+\As^2\(
128+Y^2\alpha_8(-16+Y(4+5Y)\alpha_8)\notag\\
&~~~~~~~~~~~
+4Y\alpha_8\As^2(8+Y(1+Y)\alpha_8+Y\alpha_8\As^2)
\)\].
\end{align}
The scalar sector further degenerates when $\alpha_1=1/Y$.
\\

\section{Case C : General theories with $f=0$}
\label{appendix-CaseC}

In this appendix, we 
investigate a remaining subclass of degenerate theories where
 the Einstein-Hilbert term is absent, {\it i.e.,} $f=0$. 
 In this case, 
the first condition $D_0(Y)=0$ is automatically satisfied since $Q$ vanishes with $f=f_Y=0$,
 as can be seen in (\ref{Qdef}). 
Then, we have 
\ba
0 
&=& {1\over 16} \alpha_1 Y
\(2\alpha_1+Y(\alpha_4+\alpha_8)-\beta\)\nonumber\\
&&\times\[4 \alpha _1 \left(\alpha _1+4 \alpha _2\right)
+4 \(3 \alpha _2 \alpha _4+\alpha _1\left(\alpha _3+\alpha _4\right)\) Y
+\(4\left(\alpha _1+3 \alpha _2\right) \alpha _5-3 \alpha_3^2\) Y^2 \]. 
\label{branchC}
\ea
Thus, we have four branches here.
\ba
&&\alpha_1=0
\quad  {\rm or}\quad
\alpha_4={\beta-2\alpha_1-Y\alpha_8 \over Y}\quad{\rm or}\quad\notag\\
&&
\alpha_4=\frac{3Y^2\alpha_3^2-4\alpha_1(\alpha_1+4\alpha_2)-4Y\alpha_1\alpha_3}{4Y(\alpha_1+3\alpha_2)}-Y\alpha_5
\quad  {\rm or}\quad
\alpha_1=-3\alpha_2.
\ea
Here, we assumed $\alpha_1+3\alpha_2 \neq 0$ in the third branch.
In the first branch, we automatically have $D_1(Y)=0$, therefore\\
{\bf  \underline{Case~C1} :}
\ba
&& f=\alpha_1=0. \notag\\
\ea
Note that the case C1 corresponds to the "class IIIc" in \cite{2016arXiv160208398B}.
In this case, the other determinants are all zero, $\det \M_1=\det \M_2=0$.
The second branch in (\ref{branchC}) provide us the following condition, 
\ba
0=D_1(Y)=-\frac{1}{16} \alpha _1 Y\left(\alpha _1+3 \alpha _2\right) 
\left(\alpha _8 Y-2 \beta \right){}^2.
\ea
The first solution corresponds to the case C1, then we have two solutions, \\
{\bf  \underline{Case~C2} :}
\ba
&& f=0, \quad \alpha_1=-3\alpha_2, \quad \alpha_4=\frac{6\alpha_2-Y\alpha_8+\beta}{Y},
\ea
{\bf  \underline{Case~C3} :}
\ba
&& f=0, \quad \alpha_4=-\frac{2\alpha_1+\beta}{Y}, \quad \alpha_8={2\beta \over Y}.
\ea
Note that this case includes the special case of the beyond GP theory with $f=0$ when $\alpha_1=-\alpha_2$, $\alpha_3=-\alpha_4$, and $\beta=0$.
The third branch in (\ref{branchC}) gives 
\ba
0=D_1(Y)=\frac{1}{16} \alpha _1 Y \[-4 \alpha _1 \left(\alpha _1+2
\alpha _2\right) \beta +4 \alpha _1 \alpha _3 \beta  Y
-\(3 \alpha _3^2 \beta+(\alpha _1+3 \alpha _2) (\alpha _8^2-4\alpha _5 \beta )\)Y^2 \].~~
\ea
Then, solving for $\alpha_5$, we have\\
{\bf  \underline{Case~C4} :}
\ba
&& f=0, \quad \alpha_4=-{2\alpha_1 \over Y}-{Y\alpha_8^2 \over 4\beta}, 
\quad \alpha_5=\frac{4\alpha_1(\alpha_1+2\alpha_2)-4Y \alpha_1\alpha_3 + 3 Y^2 \alpha_3^2}{4Y^2(\alpha_1+3\alpha_2)}+\frac{\alpha_8^2}{4\beta}. 
\ea
 where $\alpha_1 + 3 \alpha_2 \neq 0$ is assumed since the case $\alpha_1 + 3 \alpha_2 = 0$ is studied in the case C$2$.
Note that this case includes the beyond GP theory with $f=0$ and $\beta \neq 0$ when $\alpha_1=-\alpha_2$, $\alpha_3=-\alpha_4$, and $\alpha_8=0$.
We have an additional solution in the third branch, which is $\beta=0$. 
In this case, $D_1(Y)$ is given by 
\ba
0=D_1(Y)=-{1\over 16} Y^3 \alpha_1(\alpha_1+3\alpha_2)\alpha_8^2.
\ea
The first solution corresponds to the case C1. Since $\alpha_1+3\alpha_2\neq 0$ in this branch, 
we only have one solution, 
\\  
{\bf  \underline{Case~C5} :}
\ba
&& f=0, \quad \alpha_4=\frac{3Y^2\alpha_3^2-4\alpha_1(\alpha_1+4\alpha_2)-4Y\alpha_1\alpha_3}{4Y(\alpha_1+3\alpha_2)}-Y\alpha_5, \quad \beta=\alpha_8=0.
\ea
The last branch in (\ref{branchC}) gives
\ba
0
=-{9 \over 16} Y \alpha_2 (2\alpha_2+Y\alpha_3)^2
\(6\alpha_2-Y(\alpha_4+\alpha_8)+\beta\).
\ea
Then, the first solution $\alpha_2=0$ immediately leads to $\alpha_1=0$, which is included in the case C1. 
The second solution gives \\
{\bf  \underline{Case~C6} :}
\ba
&& f=0, \quad \alpha_1={3Y\alpha_3 \over 2}, \quad \alpha_2=-{Y\alpha_3 \over 2}.
\ea
The last solution is exactly the same one as obtained in the case C2.

\bibliographystyle{JHEPmodplain}
\bibliography{references}

\end{document}